\documentclass[10pt, aps, prd, superscriptaddress, nofootinbib, showpacs, twocolumn]{revtex4-2}
\usepackage{amsmath, amsfonts, amsthm, amssymb, mathrsfs}
\usepackage[all]{xy}
\usepackage{bm}
\usepackage{stmaryrd}
\usepackage{graphicx}
\usepackage{hyperref}
\usepackage{color}

\usepackage{pgfplots, grffile, tikz}
\pgfplotsset{compat=newest}
\usetikzlibrary{plotmarks, arrows.meta}
\usepgfplotslibrary{patchplots}

\DeclareRobustCommand{\vect}[1]{\bm{#1}}
\numberwithin{equation}{section}

\DeclareMathOperator{\tr}{tr}
\DeclareMathOperator{\Tr}{Tr}

\newcommand{\calE}{\mathcal E}
\newcommand{\calI}{\mathcal I}
\newcommand{\calM}{\mathcal M}
\newcommand{\calO}{\mathcal O}

\newcommand{\calR}{\mathcal R}

\newcommand{\bbG}{\mathbb G}

\newcommand{\bbK}{\mathbb K}

\newcommand{\bfF}{\bm{F}}
\newcommand{\bfK}{\bm{K}}
\newcommand{\bfT}{\bm{T}}

\newcommand{\lb}{\llbracket}
\newcommand{\rb}{\rrbracket}

\begin{document}

\title{The heat kernel expansion for higher order minimal and non-minimal operators}

\author{A. O. Barvinsky}
\email{barvin@td.lpi.ru}
\affiliation{Theory Department, Lebedev Physics Institute, Leninsky Prospect 53, Moscow 119991, Russia}
\affiliation{Institute for Theoretical and Mathematical Physics, Lomonosov Moscow State
University, Leninskie Gory, GSP-1, Moscow, 119991, Russia}

\author{W. Wachowski}
\email{vladvakh@gmail.com}
\affiliation{Theory Department, Lebedev Physics Institute, Leninsky Prospect 53, Moscow 119991, Russia}

\begin{abstract}
We build a systematic calculational method for the covariant expansion of the two-point heat kernel $\hat K(\tau|x,x')$ for generic minimal and non-minimal differential operators of any order. This is the expansion in powers of dimensional background field objects---the coefficients of the operator and the corresponding spacetime and vector bundle curvatures, suitable in renormalization and effective field theory applications. For minimal operators whose principal symbol is given by an arbitrary power of the covariant Laplacian $(-\Box)^M$, $M>1$, this result generalizes the well-known Schwinger--DeWitt (or Seeley--Gilkey) expansion to the infinite series of positive and negative fractional powers of the proper time $\tau^{1/M}$, weighted by the generalized exponential functions of the dimensionless argument $-\sigma(x,x')/2\tau^{1/M}$ depending on the Synge world function $\sigma(x,x')$. The coefficients of this series are determined by the chain of auxiliary differential operators acting on the two-point parallel transport tensor, which in their turn follow from the solution of special recursive equations. The derivation of these operators and their recursive equations are based on the covariant Fourier transform in curved spacetime. The series of negative fractional powers of $\tau$ vanishes in the coincidence limit $x'=x$, which makes the proposed method consistent with the heat kernel theory of Seeley--Gilkey and generalizes it beyond the heat kernel diagonal in the form of the asymptotic expansion in the domain $\sigma(x,x')\ll\tau^{1/M}$, $\tau\to 0$. Consistency of the method is also checked by verification of known results for the minimal second-order operators and their extension to the generic fourth-order operator. Possible improvement of the suggested Fourier transform approach to the noncommutative algebra of the covariant $\Box$ operator in the method of universal functional traces is also briefly discussed.
\end{abstract}

\maketitle

\section{Introduction}

The purpose of this paper is to work out efficient method of calculating the heat kernel of various differential operators in curved spacetime. It is needless to say that the heat kernel is a powerful tool in mathematical physics, which underlies numerous applications in quantum field theory, theory of gauge fields and quantum gravity, theory of stochastic processes and condensed matter systems. It is especially important in the renormalization theory of quantum gravitational models where, when combined with the background field method, it allows one to express in a closed form the Green function of the theory and its one-loop effective action. Moreover, the Schwinger--DeWitt \cite{Schwinger, DeWitt, Barvinsky85, Scholarpedia, Avram00} or Seeley--Gilkey \cite{Seeley, Gilkey1975, Gilkey1979, Vassil03} heat kernel method enables explicit calculation of local UV divergences not as their expansion in powers of field perturbations on a simple symmetric background, but as full nonlinear counterterms---local nonlinear functionals of the generic background field. Starting with the pioneering application in quantum Einstein theory \cite{tHooftVeltman}, this method proved to be very efficient and now underlies majority of results on renormalization of (super)gravitational models.

The basic tool of this method is the heat kernel expansion in powers of the proper time (or Schwinger parameter). The coefficients of this expansion---the so-called HaMiDeW (Hadamard--Minakshisundaram--DeWitt) \cite{Gibbons} or Seeley--Gilkey coefficients---represent the sequence of objects of growing dimensionality directly related to the operators of field and gradient expansion in effective field theory. Thus in the lowest orders corresponding to essential and marginal operators they contain the information about UV renormalization properties of the theory  and also, within a needed accuracy below the cutoff, incorporate all quantum corrections.

A well known difficulty with the extension of this method to a wider class of field models is that it is directly applicable only to the theories with the inverse propagator which is a second-order {\em minimal} differential operator
\begin{equation}
\hat F(\nabla) =  -\Box\,\hat 1 + \hat P,
\end{equation}
whose derivatives form a covariant Laplacian $\Box = g^{ab}\nabla_a\nabla_b$ defined in a curved $d$-dimensional spacetime with the Riemannian metric $g_{ab}(x)$.\footnote{In what follows we work in the Euclidean space version of the theory related to the physical Lorentzian signature spacetime by Wick rotation, so that $-\Box$ is positive-definite under appropriate boundary conditions.} Here the covariant derivatives (and the operator as a whole) are acting on a generic set of fields $\varphi(x)=\varphi^A(x)$ with spin-tensor labels $A$ of arbitrary nature, and the hat denotes matrices in the vector space of $\varphi^A$, in particular $\hat 1=\delta^A_B$ denoting the corresponding unit matrix. Only in this particular case there exists the Schwinger--DeWitt ansatz for its heat kernel $\hat K(\tau|x,x') = e^{-\tau \hat F(\nabla)}\delta(x,x')$ of the form
\begin{multline} \label{HeatKernel0}
\hat K(\tau|x,x') = \frac{\Delta^{1/2}(x,x')}{(4\pi\tau)^{d/2}}\, g^{1/2}(x') \\
\times \exp\left(-\frac{\sigma(x,x')}{2\tau}\right) \sum\limits_{m=0}^\infty \tau^{m}\,\hat a_m(x,x').
\end{multline}
Here $\sigma(x,x')$ is the Synge world function---one half of the square of the geodetic distance between the points $x$ and $x'$, $\Delta(x,x')$ is the dedensitized Pauli--Van Vleck--Morette determinant \eqref{Van_Vleck}, $g^{1/2}(x')$ expresses the fact that the heat kernel as well as the delta function are densities of weight one in the second argument and $\hat a_m(x,x')$ are well-known HaMiDeW or Schwinger--DeWitt coefficients. These coefficients satisfy recurrent differential equations which can be successively solved for $\hat a_m(x,x')$ in the form of covariant Taylor series in powers of $\sigma^{a'} = \nabla^{a'}\!\sigma$---the vector tangential at the point $x'$ to the geodetic connecting $x$ and $x'$ (generalizing the notion of the flat space vector $x' - x$ to curved spacetime). The coefficients of this Taylor expansion are local functions of spacetime metric, its curvature and background fields, and thus provide all the goals of perturbative UV renormalization of local field models and their effective field theory expansion.

Unfortunately, this powerful technique does not directly apply to higher order operators or to {\em non-minimal} operators of any order, when their highest derivative term is not a multiple of a unit matrix in the space of fields. Examples of this happen already in the simplest case of the electromagnetic field treated in a generic Lorenz covariant gauge with $\hat F(\nabla)\equiv F^a_b(\nabla) = \Box\,\delta^a_b + \lambda\nabla^a\nabla_b + \ldots$, for a wide class of such gauges in Einstein theory \cite{Barvinsky85} and in higher-derivative gravity theories \cite{Fradkin1982, AvramBarvinsky}. There exist indirect methods which reduce the problems with higher-derivative and non-minimal operators to those of the minimal second-order one --- operator factorization into the product of second-order ones and the method of the so-called universal functional traces \cite{JackOsborn,Barvinsky85}. The latter is based on the calculation of the sequence of coincidence limits of the following two-point kernels
\begin{equation} \label{UFT}
\left.\nabla_{a_1}...\nabla_{a_n} \frac1{(-\Box)^m}\,\delta(x,x')\,\right|_{x=x'},
\end{equation}
which are calculable by the Schwinger--DeWitt technique of the above type, because the $m$-th power of the inverse Laplacian is easily generated via the proper time integral of the heat kernel of the minimal operator $\Box$.\footnote{The essence of this method can be demonstrated on the example, say, of the higher derivative theory with the inverse propagator $(-\Box)^N+P(\nabla)$. By expanding the one-loop functional determinant, $\Tr\ln\big((-\Box)^N+P(\nabla)\big)=N\,\Tr\ln\Box + \Tr\ln\big(1+P(\nabla)/(-\Box)^N\big)$, in powers of the nonlocal term $P(\nabla)/(-\Box)^N$ and commuting to the left the powers of $P(\nabla)$ and to the right---the inverse powers of $\Box$, one finds that the result will be given by the infinite series of terms (\ref{UFT}) multiplied by tensors of ever growing dimensionality.} This method turned out to be very efficient in the calculation of beta functions of the $(3+1)$-dimensional Ho\v{r}ava gravity model \cite{Barvinsky2021ubv}---the task which is apparently impossible to accomplish within a conventional momentum space diagrammatic technique on flat space background (because of the necessity to calculate the contributions of hundreds of thousands of relevant Feynman graphs).

The efficiency of these indirect methods, the use of which is always contingent on concrete peculiarities of the model, does not retract the necessity of the general method applicable to a widest possible scope of problems. Moreover, indirect methods usually allow one to calculate only the heat kernel diagonal $\hat K(\tau|x,x)$ or the spacetime integral of this coincidence limit, which is usually everything one needs in the one-loop approximation of quantum field theory. However, one might need it beyond this limit to obtain, for example, the two-point Green function and, what is even more important, to apply it in multi-loop orders. In addition, even the calculation of the expansion  (\ref{HeatKernel0}) requires to operate with $x'\neq x$, because the recurrent equations for $\hat a_m(x,x')$ fully involve their two-point functions rather than $\hat a_m(x,x)$. Extension beyond the diagonal $x'=x$ brings to life additional difficulty which was briefly discussed in \cite{Wach2}.

The series (\ref{HeatKernel0}) can be considered as an expansion in powers of background field dimensionality because in the coincidence limit every HaMiDeW-coefficient has the structure of the sum of monomials in spacetime curvatures of dimension two in units of inverse length $l$, $\calR \sim 1/l^2$, collectively denoted by $\calR$, and their covariant derivatives
\begin{equation}
\hat a_m(x,x) \propto \sum\nabla^n \calR^k, \qquad n+2k=2m.
\end{equation}
Denoting the dimension of such terms $\calO \sim 1/l^n$ by $\dim\calO = n$, one has for the second order operator $\hat F(\nabla)$ and its proper time parameter $\tau$
\begin{equation}
\dim\left(\nabla^n \calR^k\right) = n+2k,\qquad
\dim\tau = -2,
\end{equation}
so that every term in the expansion (\ref{HeatKernel0}) is dimensionless, and the power of $\tau$ can be regarded as grading the dimension of the corresponding coefficient $\hat a_m$.\footnote{With two-point coefficients $\hat a_m(x,x')$ the situation is more complicated because they also contain dimensional $\sigma^a(x,x')$, $\dim\sigma^a = -1$, but within the covariant Taylor expansion $\hat a_m(x,x') \sim \sum(-1)^n\sigma^{a_1}...\sigma^{a_n} (\nabla_{a_1}...\nabla_{a_n}a_m\,)_{x=x'}/n!$, and the dimension of $\sigma^a$'s is canceled by extra derivatives, so that $\dim a_m = 2m$.}

For the coincidence limit $x'=x$ the tensor structures $\nabla^n \calR^k$ and $\tau$ are everything that enters the formalism, the dimensionality of $\tau$-powers compensating in each term of the series the dimensionality of $\nabla^n \calR^k$. Nothing else can appear for dimensional reasons, and the expansion indeed runs in powers of the background dimensionality. However, for separated points the additional dimensionless structure
\begin{equation}
\frac{\sigma^{a'}(x',x)}{\tau^{1/N}},  \label{argument}
\end{equation}
with $N=2$---the order of the operator $\hat F$---enters the game. Its  appearance cannot be prohibited by any dimensional considerations, so that each tensor structure $\nabla^n \calR^k$  can enter the formalism with arbitrarily high power of $\sigma^{a'}/\tau^{1/2}$. So a priori one cannot look for the heat kernel expansion ansatz in positive powers of the proper time. Remarkably, for minimal second order operators infinite series of negative powers of $\tau$ gets resummed into the Gaussian function of the argument (\ref{argument}), which is regular at $x'=x$ ($\sigma(x,x')=0$) but has essential singularity at $\tau=0$. The fact that it factorizes in (\ref{HeatKernel0}) as an overall multiplier is very reminiscent of the semiclassical approximation with $\tau\to 0$ playing the role of the Planck constant $\hbar$ \cite{Wach2}.

One could have expected that the same resummation in powers of (\ref{argument}) would be possible for {\em minimal higher order}\footnote{ A typical example of higher-order minimal operators are the so-called ``conformally covariant differential operators'', such as the 4th order Paneitz operator
$$
\Delta_4 = \Box^2 + 2R^{ab}\nabla_a\nabla_b - \tfrac{2}{3}R\Box + \tfrac{1}{3}(\nabla^aR)\nabla_a
$$
and its higher-order analogs. There is a vast literature devoted to the study of such operators, see, for example, \cite{Paneitz, Branson, Erdmenger}} operators of generic order $N=2M>2$
\begin{equation}
\hat F(\nabla) = (-\Box)^M\hat 1 + \hat P(\nabla),
\end{equation}
where $\hat P(\nabla)$ is its lower derivative part (note that now $\dim\tau = -N$). However, the attempt of such a resummation in terms of the {\em generalized exponential functions (GEF)} fails \cite{Wach2}, which is apparently related to the failure of the relevant semiclassical approximation \cite{MaslovFedoriuk} in the vicinity of $x'=x$.\footnote{As discussed in \cite{Wach2}, asymptotic expansion obtained by the saddle-point approximation at $\tau\to 0$ works well as a generalized function, satisfying in particular the required initial condition $K(0|x,x')=\delta(x,x')$, but this expansion for $M>1$ is not homogeneous for $x'\to x$ and, therefore, gives singular results for the physically important coincidence limit $K(\tau|x,x)$. This makes it impossible to apply in this coincidence limit the semiclassical approximation of \cite{MaslovFedoriuk}.} This also invalidates the attempt, undertaken in \cite{Fulling} to build the recurrent relations for the generalized HaMiDeW-coefficients for fourth-order operators, because of the infinite series of negative powers in $\tau$.

Thus, the goal of this paper is to try circumventing these difficulties and develop a workable calculational method for generic higher-derivative operators. Rather than merely formulating general statements on the structure of asymptotic expansion for the functional trace of the heat kernel, well known from mathematics literature \cite{Seeley, Gilkey1975, Gilkey1979, Gilkey2003, Gilkey1980, GilkeyFegan, GilkeyBransonFulling}, we will derive concrete algorithms for the expansion of the two-point heat kernel with separate points. In view of the discussion above, for $x\neq x'$ it will not be an asymptotic expansion in $\tau\to 0$, because of the presence of infinite series with negative powers of the proper time. Rather it will be an expansion in the operators of positive dimension, the spacetime curvature, background fields and their derivatives---what is just needed in UV renormalization and effective field theory. Concrete algorithms for this expansion will involve recurrent procedure for the calculation of what we call {\em generalized HaMiDeW coefficients}---the coefficients of the series in $\tau$, weighted by the sequence of GEFs that were introduced in \cite{Wach2}. For generic higher-derivative operators these functions replace the overall usual exponential function of the Schwinger--DeWitt series (\ref{HeatKernel0}). We will show that in the coincidence limit $x'=x$ all generalized HaMiDeW-coefficients of negative powers of $\tau$ vanish and, therefore, result in the conventional asymptotic $\tau\to 0$ expansion for the heat kernel diagonal.

The method of derivation of all these results, is the generalized Fourier transform in curved spacetime. Basically, it consists in the replacement of the usual momentum space integrals in Cartesian coordinates by the integrals over the momentum dual to the distinguished coordinate variable---the vector $\sigma^{a'}(x,x')$ tangential to the geodetic interpolating between the points $x$ and $x'$. In fact, the Fourier integral method underlies the original statements of spectral geometry of (pseudo)differential operators of \cite{Seeley, Gilkey1975}. { Its covariant version was firstly introduced by H.~Widom \cite{Widom1, Widom2, Widom3} and then it was used in the works by Gusynin {\it et al} \cite{Gusynin1989, Gusynin1990, Gusynin1991, GusyninGorbar, GusyninGorbarRomankov, Gorbar}} addressing various problems with second-order non-minimal and minimal higher-order operators. Here we systematically use this method starting with generic non-minimal operator of $N$-th order, then push forward the resulting formalism to the case of minimal operators of arbitrary even order $N=2M$. Then we show how the case of non-minimal operators, which satisfy the so-called causality condition \cite{Barvinsky85}, can be fully reduced to the minimal case.

The paper is organized as follows. Main results for two-point heat kernel are listed in Sect. \ref{Main}. Their derivation is given in Sects. \ref{Fourier} and \ref{MinimalOperators}, along with a special emphasis on heat kernel diagonal elements. In Sect. \ref{Examples} we recover known results of the Schwinger-DeWitt technique for minimal second order operators. In Sect. \ref{4thOrder} these results are extended to generic minimal fourth-order operators. Sect. \ref{PowerMin} is devoted to relations between the heat kernels of differential operators and their powers, including non-integer powers corresponding to pseudo-differential case. Sect. \ref{Conclusion} contains the discussion of the suggested technique along with its computational limitations, which might perhaps be circumvented within non-commutative algebra of the covariant $\Box$ operator in the method of universal functional traces of \cite{Barvinsky85}.

\section{Summary of main results \label{Main}}

We begin with the definition of the covariant matrix-valued differential operator $\hat F(\nabla)=F^A_B(\nabla)$ of order $N$ acting on generic set of fields $\varphi^B(x)$ in $d$-dimensional spacetime
\begin{equation}\label{GO}
\hat F(\nabla) = \sum\limits_{k=0}^N \hat F_k^{a_1\ldots a_k}(x) \nabla_{a_1}\ldots\nabla_{a_k},
\end{equation}
where $\hat F_k^{a_1\ldots a_k}(x)$ are its coefficients. Since commutation of covariant derivatives leads to a decrease in the order, one can consider these coefficients $\hat F_k(x)$ as completely symmetric tensors. Hereinafter, we will systematically omit convolutions over repeated multiple indices and use the following abbreviated notation
\begin{equation} \label{AbbrInd}
\hat F_k * \nabla^k = \hat F_k^{a_1\ldots a_k} \nabla_{a_1}\ldots\nabla_{a_k}.
\end{equation}
Thus,
\begin{equation}
\hat F(\nabla) = \sum\limits_{k=0}^N \hat F_k(x) * \nabla^k.
\end{equation}

The main result for the heat kernel $\hat K(\tau| x,x')$ of a generic operator (\ref{GO}) can be formulated in terms of the following ingredients of the geodetic interpolating between the points $x$ and $x'$---the parallel transport tensor $\hat\calI(x,x')$, the vectors $\sigma^a(x,x')$ and $\sigma^{a'}(x,x')$  tangent to this geodetic respectively at $x$ and $x'$ and the sequence of special two-point tensor and matrix valued differential operators $\hat T_{k,l}(\nabla)=\hat T_{k,l}^{a'_1\ldots a'_l}(x,x'|\nabla)$ which will be determined by the series of recurrent equations in Sect.\ref{Fourier}. The resulting heat kernel expansion reads
\begin{align} \label{bbKRazl}
&\hat K(\tau| x,x') = \sum\limits_{k=0}^\infty \tau^n\!\!\!\! \sum\limits_{\;\;\;0\le l\le L_n}\!\!\!\! \hat S_l(\tau) * \hat T_{n,l}(\nabla)\, \hat\calI(x,x'),\\
&L_n(N) = \Big(N-\frac{1}{2}\,\Big)\,n, \label{Ln}
\end{align}
where $\hat S_l(\tau)=\hat S_{l,b'_1\ldots b'_l}(\tau|x,x')$ are the two-point matrix-valued tensors of $l$-th rank at $x'$ given by the following momentum integrals
\begin{multline} \label{IntL3}
\hat S_{l,b'_1\ldots b'_l}(\tau| x,x') = \big(\det\bar\sigma^a_{b'}\big)\, \bar\sigma^{a_1}_{b'_1}\ldots\bar\sigma^{a_l}_{b'_l} \\
\times \int \frac{d^dp}{(2\pi)^d} (ip_{a_1})\ldots (ip_{a_l})\, \exp\left(-\tau\hat F_N(x, \bm{p}) + ip_a \sigma^a\right)
\end{multline}
with $\hat F_N(x, \bm{p})$---the matrix-valued principal symbol of the operator $\hat F(\nabla)$
\begin{equation}
\hat F_N(x, \bm{p}) = \hat F_N^{a_1\ldots a_N}(x) \times (ip_{a_1})\ldots\times(ip_{a_N}).   \label{symbol}
\end{equation}

Here the tangent geodetic vectors $\sigma^a(x,x')$, $\sigma^{a'}(x,x')$ and the vector bundle parallel transport tensor $\hat\calI(x,x')$ satisfy the equations
\begin{align}
&\sigma^a\nabla_a\hat\calI = 0,
\qquad \big[\,\hat\calI\, \big] = \hat 1,   \label{I_eq}\\
&\sigma^b\nabla_b\sigma^{a'}=\sigma^{a'}, \;\; \sigma^{b'}\nabla_{b'}\sigma^a=\sigma^a,\;\;
\big[\,\sigma^a \big]=\big[\,\sigma^{a'}\big]=0, \label{sigma_eq}
\end{align}
and in their turn determine the matrix $\sigma^{b'}_a=\sigma^{b'}_a(x,x')$ and its inverse $\bar\sigma^a_{b'}=\bar\sigma^a_{b'}(x,x')$,
\begin{equation} \label{bitensor}
\sigma^{b'}_a = \nabla_a\sigma^{b'}, \quad \bar\sigma^a_{b'}\sigma^{b'}_c=\delta^a_c,
\end{equation}
for any two-point function squared brackets are its coincidence limits
\begin{equation}
[f(x,x')] = f(x, x).
\end{equation}

It is important that, although the expansion (\ref{bbKRazl}) looks as the expansion in $\tau\to 0$, in fact it is not a small time expansion of the heat kernel, because the calculation of the integrals (\ref{IntL3}) can lead to arbitrarily high negative powers of $\tau$. Rather, this is the expansion in powers of the background dimensionality---what is exactly needed in renormalization and effective field theory. This follows from the fact that the differential operators $\hat T_{n,l}(\nabla)$ have the dimensionalities,
\begin{equation}
\dim\,\hat T_{n,l}(\nabla) = nN-l \geq \frac{n}2,   \label{dim_T}
\end{equation}
monotonically growing with $n$ in view of the upper bound (\ref{Ln}) on the summation index $l$ in (\ref{bbKRazl}).

Note that a generic operator (\ref{GO}) does not assume the existence of the spacetime metric and involves only the vector bundle connection in $\nabla_a\varphi=(\partial_a+\hat\varGamma_a)\,\varphi$ and the affine connection which without loss of generality we assume to be symmetric. Therefore, the tangent geodetic vectors at this stage are not the gradients of the world function $\sigma(x,x')$ which can be introduced only along with the metric structure.

The metric structure emerges together with the definition of the minimal differential operators. Their highest derivative term is defined as the power of the covariant Laplacian $\Box=g^{ab}\nabla_a\nabla_b$ times the unit matrix $\hat 1$ in the vector space of the fields $\varphi$. Thus, a minimal differential operator of general even order $N=2M$ is defined as
\begin{eqnarray}
&&\hat F(\nabla) = (-\Box)^M\,\hat 1 + \hat P(\nabla),\label{Min_O}\\
&&\hat P(\nabla)=
\sum\limits_{k=0}^{2M-1}\hat P_k(x)*\nabla^k, \label{Min_O1}
\end{eqnarray}
where $\hat P(\nabla)$ is the lower derivative part---an arbitrary differential operator of the order $2M-1$.

As long as the spacetime metric $g_{ab}$ now enters the formalism, we will assume that covariant derivatives conserve it, $\nabla_a g_{bc}=0$, and the tangent geodetic vectors $\sigma^{a'}=\sigma^{a'}(x',x)$ and $\sigma^a=\sigma^{a}(x,x')$,
\begin{equation}
\sigma^a=\nabla^a\sigma,\quad
\sigma^{a'}=\nabla^{a'}\sigma, \label{sigma_grad}
\end{equation}
become the gradients of the world function $\sigma=\sigma(x,x')$---the squared geodetic distance between $x$ and $x'$, which satisfies with respect to its both arguments the equations
\begin{equation}
\sigma = \frac12\,\sigma^a\sigma_a  = \frac12\,\sigma^{a'}\sigma_{a'}. \label{sigma_eq1}
\end{equation}

The principal symbol (\ref{symbol}) of a minimal operator simplifies to
\begin{equation} \label{symbol1}
\hat F_N(x,\bm{p})=\big(g^{ab}(x)\,p_a p_b\big)^M\, \hat 1\equiv p^{2M}\,\hat 1,
\end{equation}
and the set of integrals (\ref{IntL3}) expresses via a single basic integral in terms of the {\em generalized exponential function (GEF)} $\calE_{M, d/2}(z)$ of the argument $z=-\sigma/2\tau^{1/M}=-\sigma^a\sigma_a/4\tau^{1/M}$,
\begin{multline} \label{calEInt}
\int \frac{d^dp}{(2\pi)^d} \exp\left(-\tau p^{2M} + ip_a \sigma^a\right) \\
= \frac{g^{1/2}(x)}{\left(4\pi\tau^{1/M}\right)^{d/2}}\, \calE_{M, d/2}\left(-\frac{\sigma}{2\tau^{1/M}}\right).
\end{multline}
This function was introduced in \cite{Wach2}, and its main properties are presented in Appendix \ref{GEF}. Via the differentiation with respect to $\sigma^b$ it generates a sequence of new fully symmetric tensor coefficients
\begin{equation}
S_{p,l}=S_{p,l,\,a'_1\ldots a'_l}(x,x'). \label{Spl}
\end{equation}
These are the polynomials in powers of $\sigma_{a'}$ and the factors of the form $\gamma_{a'b'} = \bar\sigma_{a'}^c g_{cd} \bar\sigma^d_{b'}$, which arise according to the equation
\begin{multline} \label{IntL4}
\Big(\prod\limits_{j=1}^l\bar\sigma^{b_j}_{a'_j} \frac{\partial}{\partial\sigma^{b_j}}\Big)\, \calE_{M,\frac{d}{2}} \left(-\frac{\sigma^b\sigma_b}{4\tau^{1/M}}\right)\\
=  \sum\limits_{p\geq\frac{l}{2}}^l \frac{S_{p,l,a'_1\ldots a'_l}
}{\left(-2\,\tau^{1/M}\right)^p}\,\calE_{M,\frac{d}{2}+p} \left(-\frac{\sigma}{2\,\tau^{1/M}}\right),
\end{multline}
due to the main property (\ref{DiffRule}) of $\calE_{\nu, \mu}(z)$, $\partial_z\calE_{\nu,\mu}(z)=\calE_{\nu,\mu+1}(z)$.

Main result for the heat kernel of a generic minimal operator of any order $2M$ is then as follows. Its expansion in fractional powers of proper time reads as
\begin{multline} \label{MainExpansion}
\hat K(\tau| x,x') =  \frac{\Delta^{-1}(x,x')}{(4\pi\tau^{1/M})^{d/2}}\,g^{1/2}(x')
\sum\limits_{m=-\infty}^\infty \tau^{m/M} \\
\times \sum\limits_{\,n \ge N_m}^\infty \calE_{M,\frac{d}2+M n-m}\left(-\frac{\sigma}{2\tau^{1/M}}\right)\, \hat b_{m,n}(x,x').
\end{multline}
where the lower bound on the summation index $n$ in each $m$-th order of $\tau^{1/M}$ is
\begin{eqnarray} \label{N_m}
N_m(M) = \begin{cases}
\,\frac{m}{M}, & m>0, \\
&\\
\,\frac{2|m|}{2M-1}, & m<0,
\end{cases}
\end{eqnarray}
$\Delta(x,x')$ is the Pauli--Van Vleck--Morette determinant \eqref{Van_Vleck} and $\hat b_{m,n}(x,x')$ are {\em the generalized HaMiDeW coefficients} given by the following finite sums in terms of the above operators $\hat T_{n,l}(\nabla)$ and tensor coefficient functions $S_{p,l}(x,x')$,
\begin{multline}\label{bmn}
\hat b_{m,n}(x,x') = \frac{1}{(-2)^{Mn-m}} \\
\times \sum\limits_{l=Mn-m}^{\lfloor L_{m,n}\rfloor} S_{Mn-m,l}(x,x') * \hat T_{n,l}(\nabla)\,\hat\calI(x,x').
\end{multline}
Here the upper bound on the summation index $l$ is the integer part $\lfloor L_{m,n}\rfloor$ of
\begin{eqnarray} \label{Rnm}
L_{m,n}&=&2Mn-{\rm max}\left\{\,2m, \frac{n}2\,\right\}.
\end{eqnarray}

This is again effectively the expansion in background dimensionality, because as it follows from (\ref{dim_T}) the dimensionalities of $\hat b_{m,n}(x,x')$ are always positive and grow with $m$ and $n$,
\begin{equation} \label{dim_b}
\dim\, \hat b_{m,n}\geq\min_{\{\,l\,\}\vphantom{L^L}} \big(\dim\,\hat T_{n,l}\big)=\max\left\{\,2m, \frac{n}2\,\right\}.
\end{equation}

Below we go over to the derivation of the above results along with the formulation of recurrent equations for the operators $\hat T_{k,l}^{a'_1\ldots a'_l} (x,x'|\nabla)$. This will be followed by verification of the consistency of the obtained formalism with known properties of the Schwinger--DeWitt technique.

\section{General non-minimal operators} \label{Fourier}

We define the heat kernel $\hat K(\tau| x, x')$ of $\hat F$ as the kernel of the operator $e^{-\tau\hat F}$, i.e. as the result of acting by this operator on the two-point delta function $\hat\delta(x,x')=\delta^A_{B'}(x,x')$ in the space of fields $\varphi=\varphi^A(x)$,
\begin{equation}
\hat K(\tau| x, x') = e^{-\tau\hat F}\,\hat\delta(x,x')
\end{equation}
The delta function is itself defined by the convolution with $\varphi(x)$
\begin{equation}
\varphi(x)=\int d^dx'\,\hat\delta(x,x')\,\varphi(x')=
\int d^dx'\,\hat\delta^A_{B'}(x,x')\,\varphi^{B'}(x'),
\end{equation}
from which it follows that $\hat\delta(x,x')$ is the density of unit weight in $x'$ and has zero weight in the first argument.\footnote{In view of the point-like support of the delta-function, it can be written down as $\hat 1\,\delta(x,x')=\delta^A_{B'}\delta(x,x')$, but the distinction between the indices $A$ and $B'$ associated respectively with the first and the second spacetime points should be clearly kept, especially when the delta function is acted upon by the covariant derivatives in the vector bundle of $\varphi^A(x)$.}

\subsection{The generalized Fourier transform}

The covariant Fourier transform method requires to consider the ``plane-wave'' functions with the momentum vector $k_{b'}$,
\begin{equation}
\exp\left(ik_{b'} \sigma^{b'}(x,x')\right), \label{plane}
\end{equation}
which are based on the coordinate vector $\sigma^{b'}(x,x')$---the analogue of the vector $x'-x$ in flat spacetime. This is the vector tangential at the point $x'$ to the geodetic curve interpolating between the points $x'$ and $x$ and satisfying the equation and initial conditions (\ref{sigma_eq}).\footnote{We remind that at this point we do not need a metric $g_{ab}$ on the spacetime manifold, which is not at all encoded in the generic differential operators of the form (\ref{GO}). The metric structure and associated Synge world function $\sigma(x,x')$ appears later for minimal differential operators, in which case the vector $\sigma^{b'}(x,x')$ expresses via $\sigma(x,x')$ as $\sigma^{b'}(x,x')=\nabla^{b'}\sigma(x,x')$.} Note that with this definition the plane waves (\ref{plane}) are scalars at the point $x$---bear in mind that $\sigma^{b'}(x,x')$ is here the scalar as a function of $x$ and the vector cotangent to spacetime at the second point $x'$.

The generalized Fourier transform in curved spacetime is based on the covariant integral representation for the delta-function. For the delta-function acting on a scalar it follows from an obvious relation
\begin{equation} \label{scalar_delta}
\delta(x,x')=\delta\big(\sigma^{a'}(x,x')\big),
\end{equation}
justified by the fact that $\big[\det\big(-\sigma^{a'}_b\big)\big] = 1$, and the subsequent representation of $\delta\big(\sigma^{a'}\big)$ in the form of the Fourier integral over the momentum space of vectors $k_{a'}$ cotangent to the spacetime at the point $x'$.

The generalization of this representation to generic vector bundle fields by a simple multiplication with $\hat 1$ will not be consistent for the following reason. In what follows we will have to work out a covariant formalism for the integrand of the corresponding momentum space integral, but the multiplication of this integrand by $\hat 1$ will not convert this object to a bi-tensor quantity with correct transformation law of its both arguments $x$ and $x'$. Thus, the spacetime covariant definition of the delta-function acting in the generic vector bundle of spin-tensor fields should read
\begin{equation}
\hat\delta(x,x') = \int \frac{d^dk}{(2\pi)^d} \exp\left(ik_{a'}\sigma^{a'}\right)\,\hat\calI(x, x'), \label{DeltaInt}
\end{equation}
where $\hat\calI(x, x')=\calI^A_{B'}(x, x')$ is some two-point spin-tensor quantity having correct transformation properties with respect to the multindex $A$ associated with $x$ and the index $B'$ associated with $x'$. The only restriction on the choice of $\hat\calI(x, x')$ is the requirement that it should reduce to the unit matrix at $x'=x$. Otherwise it can be rather arbitrary, and can generate different representations of one and the same quantity.

The simplest choice of $\hat\calI(x, x')$, one can figure out, is apparently the parallel transport tensor satisfying the equations (\ref{I_eq})
which provide the parallel translation of spin-tensor objects along the geodetic connecting the points $x$ and $x'$. This operator in fact coincides with the zeroth HaMiDeW-coefficient in the Schwinger-DeWitt expansion (\ref{HeatKernel0}), $\hat\calI(x, x')=\hat a_0(x,x')$ and in mathematics literature is usually associated with Widom calculus.

Note that within the definition (\ref{DeltaInt}) the exponential factor in the integrand of the Fourier integral is the scalar of zero weight with respect to the diffeomorphisms of $x$, and {\em after} the $k_{a'}$-integration generates the scalar density of weight one at $x'$, which is in full accordance with the transformation properties of the right hand side of (\ref{scalar_delta}).\footnote{Point-like support of the delta-function fixes only its overall weight with respect to diffeomorphisms of both arguments (which is one), but the way this weight can be splitted between $x$ and $x'$ in the integrand of the Fourier integral representation depends on the choice of this representation---with another choice, say $\delta\big(\sigma^a(x,x')\big)$, the delta-function would be a zero-weight scalar with respect to $x'$, but weight one with respect to $x$.} In contrast to this, {\em prior} to this integration the exponential function in question does not have good transformation properties as a function of $x'$ because of the contraction of $\sigma^{a'}$ with the diffeomorphism inert vector $k_{a'}$, but this drawback of the formalism can be disregarded, because we will develop a fully covariant formalism (for the Fourier image of the heat kernel) entirely at the point $x$.

Now we act by the operator $e^{-\tau\hat F(\nabla)}$ upon (\ref{DeltaInt}) and commute the exponential plane-wave function with $\hat F(\nabla)$ according to the following relation
\begin{equation} \label{commutation}
\exp\big(-ik_{b'}\sigma^{b'}\big)\hat F(\nabla_a)
\exp\big(ik_{b'} \sigma^{b'}\big) =
\hat F\big(\nabla_a + ik_{b'} \sigma^{b'}_a\big),
\end{equation}
where $\sigma^{b'}_a$ is the bi-tensor (\ref{bitensor}). Therefore, the heat kernel takes the form
    \begin{align}
    &\hat K(\tau| x, x')= \int \frac{d^dk}{(2\pi)^d}
    \exp(ik_{a'} \sigma^{a'})\, \hat\bfK(\tau, \bm{k}| x,x'), \label{IntImp} \\
    &\hat\bfK(\tau, \bm{k}|x,x')=\exp\left(-\tau\,
    \hat F\big(\nabla_a
    + ik_{b'}\sigma^{b'}_a\big)\right) \hat\calI(x, x'),    \label{bfK}
\end{align}
where its Fourier image $\hat\bfK(\tau, \bm{k}| x,x')$, which we will denote by bold letters, obviously satisfies the following initial Cauchy problem
    \begin{eqnarray}
    &&\left(\partial_\tau + \hat F\big(\nabla_a + ik_{b'} \sigma^{b'}_a \big)\right) \hat\bfK(\tau, \bm{k}| x,x') = 0,\label{HeatEqFourier}\\
    &&\hat\bfK(0, \bm{k}| x,x')
    = \hat\calI(x,x').                            \label{InitCondFourier}
    \end{eqnarray}

Why would we need the solution of this problem if we already have it in a closed form (\ref{bfK}) and can explicitly expand it in powers of the exponential? Point is that this expansion will not be a needed expansion in background dimensionality $1/l$, because both (\ref{GO}) and (\ref{commutation}) have terms of zeroth order in $1/l$---see the leading term of (\ref{GO}) in derivatives with ${\rm dim}\big(\hat F_N^{a_1\ldots a_N}(x))=0$. Thus, first we have to disentangle from $\hat F(\nabla_a + ik_{b'} \sigma^{b'}_a \big)$ this $O(1/l^0)$ term, explicitly find the zeroth order solution and then develop the perturbation theory in $1/l$.

\subsection{The solution for $\hat\bfK\left(\tau, \vect{k}| x,x'\right)$}\label{X}

To do this, we first expand the operator $\hat F(\nabla_a + ik_{b'}\sigma^{b'}_a)$ in powers of momenta $\bm{k}$,
\begin{equation} \label{ExpF1}
\hat F\left(\nabla_a + ik_{b'}\sigma^{b'}_a \right) = \sum\limits_{m=0}^N (i\bm{k})^m * \lb\hat F \rb_m,
\end{equation}
where we introduced the notations for monomials in momenta $\bm{k}^m=k_{b'_1}\ldots k_{b'_m}$ and their coefficient differential operators $\lb\hat F \rb_m = \lb\hat F \rb_m^{b'_1\ldots b'_m}(\nabla)$ and also use the same abbreviation of indices contraction as in \eqref{AbbrInd}. The rule of constructing the operators $\lb\hat F \rb_m$ actually follows from the definition (\ref{ExpF1}),
\begin{equation}
\lb\hat F\rb_m = \sum\limits_{k=m}^N
\hat F_k(x) * \lb\nabla^k\rb_m,    \label{decomp1}
\end{equation}
where the action of the operation $\lb\ldots\rb_m$ on every monomial of derivatives $\nabla^k=\nabla_{a_1}\ldots\nabla_{a_k}$ consists in replacing $m$ of them by the functions of the form $\sigma^{b'}_a$. This gives the sum of $\binom{k}{m}$ terms each having extra $m$ contravariant primed indices $b'_1,\ldots, b'_m$. The original order of derivatives and inserted factors $\sigma^{b'}_a$ should be strictly retained, and the additional obvious property holds: $\lb\nabla^k\rb_m=0$ if $m>k$.

Thus, the operation $\lb\ldots\rb_m$ reduces the order of every differential operator by $m$ and adds to it extra $m$ primed indices associated with the point $x'$. In particular, for our $N$-th order operator (\ref{AbbrInd}) its $\lb\hat F\rb_N$ is not a differential operator, but just a two-point function which is a scalar at $x$ and a tensor at $x'$ with $N$ contravariant indices
\begin{equation}
\lb\hat F\rb_N\equiv \lb\hat F\rb_N^{b'_1...b'_N} = \hat F_N^{a_1\ldots a_N} \sigma^{b'_1}_{a_1} \ldots \sigma^{b'_N}_{a_N}.
\end{equation}
When contracted with $(i\bm{k})^N$ it will play the role of the principal symbol of $\hat F(\nabla)$ usually used in the Fourier analysis of differential and pseudo-differential operators.

The dimensionality of these operators is equal to their order
\begin{equation}
\dim\, \lb\hat F \rb_m = N-m.
\end{equation}
Therefore, to develop an efficient expansion in powers of the background dimensionality it would be sufficient to expand $\hat\bfK(\tau, \bm{k}| x,x')$ in powers of all $\lb\hat F \rb_m$ with $m\leq N-1$ while treating the leading term $\lb\hat F \rb_N$ exactly. This is of course possible, because $\lb\hat F \rb_N$ is not a differential operator but just a matrix, and the lowest order solution of (\ref{HeatEqFourier}) is given by the matrix-valued exponential function
\begin{equation}
\hat\bfK_0(\tau, \bm{k}) = \exp\left(-\tau (i\bm{k})^N * \lb\hat F\rb_N \right) \hat\calI.
\end{equation}

Therefore, we will look for the solution of Eqs.\eqref{HeatEqFourier}--\eqref{InitCondFourier} in the form
\begin{eqnarray} \label{frakT}
&&\hat\bfK(\tau, \bm{k}) = \exp\Big(-\tau (i\bm{k})^N * \lb\hat F\rb_N \Big)\, \hat\bfT(\nabla)\, \hat\calI,
\end{eqnarray}
where $\hat\bfT(\nabla)\equiv\hat\bfT(\nabla, \tau, \bm{k}|x,x')$ is some unknown operator to be found.

Substituting the ansatz \eqref{frakT} into the equation \eqref{HeatEqFourier} we see that the commutation of operators $\lb \hat F\rb_m$ with the exponential function $\exp\big(-\tau (i\bm{k})^N * \lb\hat F\rb_N \big)$ leads again to elongation of derivatives
    \begin{equation}
    \nabla_a \to \tilde\nabla_a = \nabla_a - \tau (i\bm{k})^N * \hat D_a(\tau,\bm{k}),
    \end{equation}
where $\hat D_a(\tau,\bm{k})$ is given in quadratures as the following generically nontrivial matrix-valued integral
    \begin{align}
    \bm{k}^N*\hat D_a(\tau,\bm{k}) &= \frac1\tau\int_0^\tau d\tau'\,
    e^{-\tau' (i\bm{k})^N * \lb\hat F\rb_N}\nonumber\\
    &\times\big(\bm{k}^N*\nabla_a\lb\hat F\rb_N\big)\,
    e^{\tau' (i\bm{k})^N * \lb\hat F\rb_N}, \label{D}
    \end{align}
which simplifies to $\nabla_a\lb\hat F\rb_N$ for a particular case of the vanishing commutator $\big[\bm{k}^N *\lb\hat F\rb_N,\bm{k}^N*\nabla_a\lb\hat F\rb_N\big]=0$.
The resulting operators can in their turn be expanded as,
\begin{equation} \label{Commut}
\lb\hat F\rb_m(\tilde\nabla) = \sum\limits_{n=0}^{N-m} \left(\tau(i\bm{k})^N\right)^n * \lb\hat F\rb_{m,n}(\tau,\bm{k}),
\end{equation}
to give a new set of operators of the order $N-m-n$ with $m+Nn$ contravariant indices $\lb\hat F\rb_{m,n}^{b'_1\ldots b'_mc'_1\ldots c'_{Nn}}$,
\begin{align}
&\lb\hat F\rb_{m,n} = \sum\limits_{k=m+n}^N \hat F_k(x) * \lb\nabla^k\rb_{m,n},                \label{Fmn}, \\
&\dim\, \lb\hat F\rb_{m,n} = N-m, \label{DimFpq}
\end{align}
where $\lb\nabla^k\rb_{m,n}$ are built by the same pattern as
$\lb\nabla^k\rb_m$---they consist of $\binom{k}{m,n}$ terms, in each of which $m$ covariant derivatives $\nabla_a$ are replaced by the functions of the form $\sigma^{b'}_a$, and $n$ covariant derivatives $\nabla_a$ are replaced by the functions of the form $-\hat D_a(\tau,\bm{k})$. It is easy to see that $\lb\hat F\rb_{m,0} = \lb\hat F\rb_m$ and $\lb\hat F\rb_{m,N-m}$ are not operators but just tensor-valued functions of $x$ and $x'$.

After making the required commutation, one finds
\begin{align}
&\left(\partial_\tau + \hat\bfF\right) \hat\bfT(\nabla, \tau, \bm{k}) = 0, \label{FrakTEq} \\
&\hat\bfT(\nabla, 0, \bm{k}) = \hat 1, \label{FrakTInitCond}
\end{align}
where the new operator $\hat\bfF(\nabla)\equiv\hat\bfF(\nabla, \tau, \bm{k}|x,x')$ reads
\begin{equation} \label{FrakFRazl}
\hat\bfF(\nabla,\tau, \bm{k}) = \sum\limits_{m=0}^{N-1} \sum\limits_{n=0}^{N-m} \tau^{n} (i\bm{k})^{m+Nn} * \lb\hat F\rb_{m,n}(\tau, \bm{k})
\end{equation}
and has a positive dimensionality $\dim\,\hat\bfF(\nabla)\geq 1$. Now, however, this ``Hamiltonian'' operator is time dependent and admits a perturbative solution in the form of a $T$-ordered exponent
\begin{multline}
\hat\bfT(\nabla,\tau, \bm{k}) = T \exp\left(-\int_0^\tau dt\, \hat\bfF(\nabla, t, \bm{k})\right)\\
=\sum\limits_{n=0}^\infty (-1)^n\int\limits_0^\tau dt_1\int\limits_0^{t_1}dt_2\cdots\!\!
\int\limits_0^{t_{n-1}}dt_n\,
\hat\bfF(t_1)\cdots\hat\bfF(t_n), \label{Texponent}
\end{multline}
each term of which has at least of $n$-th order in background dimensionality.

In principle, every order of this expansion is explicitly calculable in terms of the powers of operators $\lb\hat F\rb_{m,n}$, introduced above, and the powers of $\tau$ and $\bm{k}$. However, since the operator (\ref{FrakFRazl}) contains at least $N(N+1)/2$ terms, the number of terms in the $n$-th order of this expansion will contain $[N(N+1)/2]^n$ monomials in powers of $\lb\hat F\rb_{m,n}$, not to mention that every $\lb\hat F\rb_{m,n}$ is a sufficiently long expression growing with the operator order $N$.

Therefore, we will consider a simpler case of the vanishing commutator $\big[\bm{k}^N *\lb\hat F\rb_N,\bm{k}^N*\nabla_a\lb\hat F\rb_N\big]=0$, when $\hat D_a = \nabla_a\lb\hat F\rb_N$ and $\lb F\rb_{m,n}$ are both $(\tau,\bm{k})$-independent and, instead of explicitly handling every order of the expansion (\ref{Texponent}), look for $\hat\bfT(\tau, \bm{k})$ in the form of double series in powers of $\tau$ and $\bm{k}$. Then we try finding solvable recurrent relations for its coefficients. This double series reads
\begin{equation} \label{FrakTRazl}
\hat\bfT(\nabla,\tau,\bm{k}) = \sum\limits_{n=0}^\infty\! \sum\limits_{\;\;\;0\le l\le L_n}\! \tau^n (i\bm{k})^l * \hat T_{n,l}(\nabla),
\end{equation}
where $\hat T_{n,l}(\nabla) = \hat T_{n,l}^{b'_1\ldots b'_l}(\nabla|x,x')$ are the unknown operator valued coefficients which are $l$-th rank tensors at $x'$ and $L_n$ is some non-decreasing sequence of finite numbers.

\begin{figure}
\includegraphics{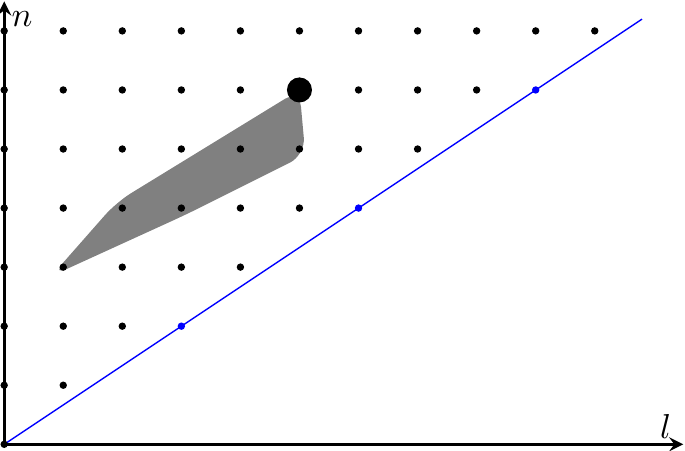}
\caption{\label{FigT} Operators $\hat T_{n,l}$ for the case $N=2$. The blue line shows lowest value of $l$, $L_n(N)=(N-1/2)n$. Every dot at $(n,l)$-position on the plane denotes the operator $\hat T_{n,l}$. Shaded domain of points shows the pattern of their causal connection to the operator in the left hand side of the recurrent equation (\ref{FrakTRecRel}) (in the example of $\hat T_{6,5}$).}
\end{figure}

The differentiation $\partial_\tau$ in \eqref{FrakTEq} decreases the power in the proper time $\tau$ by one. And the terms in the decomposition \eqref{FrakFRazl} of the operator $\hat\bfF$, on the contrary, increase the power in $\tau$ by $n$ and at the same time the power in the momentum $i\bm{k}$ by $m+Nn$. Therefore, if we substitute \eqref{FrakFRazl} and \eqref{FrakTRazl} into the equation \eqref{FrakTEq}, and then equate to zero the terms with each power $\tau^n (i\bm{k})^l$, we obtain the following system of recurrence relations for the operators $\hat T_{n,l}(\nabla)$
\begin{align}
&(n+1)\hat T_{n+1,l} = - \sum\limits_{p=0}^{N-1}
\sum\limits_{q=0}^{N-p}
\lb\hat F\rb_{p,q}\, \hat T_{n-q, l-p-Nq},\label{FrakTRecRel}\\
&\hat T_{0,0} = \hat 1, \qquad \hat T_{0,l} = 0, \quad l>0. \label{InitT}
\end{align}
This notation implies that $\hat T_{k,l} = 0$ for $k<0$ or $l<0$ and, as it should be, the composition of $(p+Nq)$-rank tensor $\lb\hat F\rb_{p,q}$ with the $(l-p-Nq)$-rank tensor $\hat T_{n-q, l-p-Nq}$ on the right hand side of this equation gives the $l$-rank tensor on the left. Initial values (\ref{InitT}) follow from the initial condition \eqref{FrakTInitCond} and thus allow one to solve these recurrent relations for all $\hat T_{n,l}$. An example of the calculation of these operators from those of the lower orders is depicted in Fig. \ref{FigT} for the case of $\hat T_{6,5}(\nabla)$ and $N=2$.

Note that the ratio of the increment of the index $l$ to the increment of the index $n$ in the transition from the right hand side of \eqref{FrakTRecRel} to its left hand side is maximal at $q=1$ and $p=N-1$
\begin{equation}
\frac{\Delta l}{\Delta n} = \frac{p+Nq}{q+1}, \qquad \max_{p,q} \frac{\Delta l}{\Delta n} = N-\frac{1}{2}.
\end{equation}
This means that the recursive calculation of $\hat T_{n,l}$, starting from a given value of $\hat T_{0,0}$, for a given $n$ involves only the values of $l$ below the bound $L_n(N)$ given by Eq.(\ref{Ln}). This specifies the finite upper limit on the summation index $l$ that we introduced in \eqref{FrakTRazl}.

Note that with the help of the relation
\begin{equation} \label{ProdOpBrak}
\lb\hat A \hat B\rb_n = \sum\limits_{k=0}^n \lb\hat A\rb_k \lb\hat B\rb_{n-k},
\end{equation}
which is valid for any two differential operators $\hat A(\nabla)$ and $\hat B(\nabla)$, one can resolve the chain of recurrence relations \eqref{FrakTRecRel} for some of the coefficients $\hat T_{k,l}$, obtaining the following closed expression
\begin{equation}
\hat T_{k,l} = \frac{1}{k!}
\left\lb(-\hat F)^k\right\rb_l, \quad l<N.  \label{special_T}
\end{equation}
In addition, the edge coefficients $\hat T_{2k,(2N-1)k}$ are also given by the explicit formula
\begin{equation}
\hat T_{2k,(2N-1)k} = \frac{1}{2^k k!}
\left(-\lb\hat F\rb_{N-1,1}\right)^k.     \label{UltraMarginal}
\end{equation}

However, an arbitrary coefficient $\hat T_{k,l}$ cannot be obtained by such simple formulas. We can only say that it can be represented as a sum of terms (with some coefficients) of the form
\begin{equation}
\lb\hat F^{k_1}\rb_{m_1,n_1}\ldots \lb\hat F^{k_p}\rb _{m_p,n_p},
\end{equation}
where
\begin{equation}
\sum\limits_{i=1}^p (k_i + n_i) = k, \qquad \sum\limits_{i=1}^p(N n_i + m_i) = l,
\end{equation}
which is consistent with the T-exponent expansion (\ref{Texponent}).

\subsection{Momentum space integration}

After we have obtained the expansion for $\hat\bfK(\nabla,\tau, \bm{k})$, it remains to integrate it over the momentum $\bm{k}$ to get the required expansion. Substituting the expressions \eqref{frakT} and \eqref{FrakTRazl} into the integral \eqref{IntImp}, we get the main result (\ref{bbKRazl})-(\ref{Ln}) for the heat kernel of a generic positive differential operator of order $N$, where $\hat S_l(\tau)=\hat S_{l,b'_1\ldots b'_l}(\tau|x,x')$ are the two-point matrix-valued tensors of $l$-th rank at $x'$ given by the following momentum integrals
\begin{equation} \label{IntL}
\hat S_l(\tau) = \int \frac{d^dk}{(2\pi)^d} (i\bm{k})^l \exp\left(-\tau (i\bm{k})^N\cdot \lb\hat F\rb_N + ik_{a'} \sigma^{a'}\right).
\end{equation}

First of all, note that the replacement of the integration momentum $k_{a'}$ belonging to the cotangent space at the point $x'$ of spacetime to the the momentum $p_a$ cotangent at the point $x$,
\begin{equation} \label{Zamena}
k_{a'} \mapsto p_a = \sigma^{b'}_a k_{b'},
\end{equation}
leads to the expression (\ref{IntL3}) for $S_l(\tau| x,x')$
with $\bar\sigma^a_{b'}$---the matrix inverse to $\sigma^{b'}_a$, \eqref{bitensor}, and with $\hat F_N(x, \bm{p})$---a usual matrix-valued principal symbol (\ref{symbol}) of $\hat F(\nabla)$. In the transition to (\ref{IntL3}) we used the equation  $\bar\sigma^a_{b'}\sigma^{b'}=\sigma^a$---the corollary of (\ref{sigma_eq}).

It is important that the expansion (\ref{bbKRazl}) is efficient from the viewpoint of effective field theory, because with the growing $k$ the background dimensionality of each term is also monotonically growing. Indeed, from the recurrence relations \eqref{FrakTRecRel} and Eq.\eqref{DimFpq}, it follows that the operators $\hat T_{n,l}(\nabla)$ have the following dimensions
\begin{equation} \label{DimTnl}
\dim\hat T_{n,l}(\nabla) = Nn - l,
\end{equation}
and in view of the upper limit \eqref{Ln} of the summation index $l$ the relation (\ref{dim_T}) holds, so that (\ref{bbKRazl}) is really the expansion in powers of spacetime curvature and dimensionful background fields.

It should be emphasized that in this counting the physical dimensionality of $\hat S_l(\tau)$, $\dim\hat S_l(\tau) = l+d$, should not be accounted for, because this factor only contributes powers of $\tau$ and the powers of the dimensionless ratio (\ref{argument}) discussed in Introduction.\footnote{Powers of $\tau$ are only grading the dimensionality of their dual operators in the expansion of the quantity of some fixed overall dimensionality. On integration over $\tau$, which generates a physical quantity like the effective action or the Green function, this parameter gets replaced by the inverse of the cutoff, $\tau\to 1/\varLambda^N$, which suppresses the contributions of higher-dimensional operators ${\cal O}_n$ of the dimension $n$, ${\cal O}_n/{\cal\varLambda}^n$, in effective field theory or inessential operators in UV renormalization. This is why the physical dimension of $\tau$ does not enter the background dimensionality.} This follows from rescaling in (\ref{IntL}) the integration momentum by $\tau^{1/N}$ and expanding the result in powers of $\sigma^{a'}/\tau^{1/N}$,
\begin{equation} \label{IntL1}
\hat S_l(\tau) = \tau^{-\frac{d+l}{N}}
\sum\limits_{p=0}^\infty \tau^{-\frac{p}{N}}\, \hat s_{l,b'_1\ldots b'_p}\, \sigma^{b'_1}\ldots
\sigma^{b'_p}.
\end{equation}
With this expression for $\hat S_l(\tau)$ it becomes obvious that (\ref{bbKRazl}) is not an expansion in $\tau$ at $\tau\to 0$, for its every $n$-th term of the dimensionality $nN-l$ carries infinitely many negative powers of the proper time. Obviously, they all vanish in the coincidence limit and do not break the validity of the effective theory expansion.

\section{Minimal operators} \label{MinimalOperators}

For the most general differential operator one cannot say much about the analytical properties of the integrals (\ref{IntL3}) except their expansion (\ref{IntL1}). In addition, the exact answer for the matrix integral (\ref{D}), $\hat D_a(\tau,\bm{k})$, with non-commutative $\bm{k}*\lb\hat F\rb_N$ and $\bm{k}*\nabla_a\lb\hat F\rb_N$ (which is needed beyond the expansion in $\lb\hat F\rb_N$ because ${\rm dim}\lb\hat F\rb_N=0$) is also hard to obtain in a closed form. For special cases a partial resummation of this expansion is, however, possible. First of all these are \emph{minimal} operators. Then we generalize this simplest case to the more general non-minimal but \emph{causal} operators of Lorentz invariant theories.

\subsection{Minimal differential operator of general even order}

For minimal operators of general even order $N=2M$, defined by Eqs.(\ref{Min_O})-(\ref{Min_O1}) their principal symbol (\ref{symbol}) simplifies to the expression (\ref{symbol1}) and tangent geodetic vectors become gradients (\ref{sigma_grad}) of the Synge world function which satisfies Eqs.(\ref{sigma_eq1}). Moreover, since the principal symbol matrix is trivial, $\lb\hat F\rb_N\propto\hat 1$, the integral (\ref{D}) yields $\hat D_a(\tau,\bm{k}) = \nabla_a\lb\hat F\rb_N$ and leads to relevant $(\tau,\bm{k})$-independent operators $\lb\hat F\rb_{m,n}$ and $\hat T_{n,l}(\nabla)$. Then the set of integrals (\ref{IntL3}) expresses via the basic integral (\ref{calEInt}) of Sect.\ref{Main} in terms of the {\em generalized exponential function (GEF)} $\calE_{M, d/2}(z)$ of the argument
\begin{equation}
z=-\frac\sigma{2\tau^{1/M}}=-\frac{\sigma^a\sigma_a}{4\tau^{1/M}}.
\end{equation}
This function, which was introduced in \cite{Wach2}, partially performs resummation of $1/\tau$ series in the heat kernel theory of higher-derivative operators, its main properties being presented in Appendix \ref{GEF}.

Bearing in mind that the powers of momenta $ip_b$ in (\ref{IntL3}) can be generated by the differentiation of this basic integral with respect to $\sigma^b$ one finds that the functions $\hat S_l(\tau)$ are given by the finite sum of terms multiple of the unit matrix---tensor densities at $x'$,
\begin{multline} \label{IntL5}
\hat S_{l,a'_1\ldots a'_l}
= \frac{\Delta^{-1}(x,x')\,g^{1/2}(x')}{\left(4\pi\tau^{1/M}\right)^{d/2}}\\
\times \sum\limits_{r\geq\frac{l}{2}}^l \frac{S_{r,l,a'_1\ldots a'_l}} {\left(-2\,\tau^{1/M}\right)^r}\,\calE_{M,\frac{d}{2}+r} \left(-{\sigma/2\,\tau^{1/M}}\right)
\,\hat 1,
\end{multline}
where the new coefficients $\hat S_{r,l,a'_1\ldots a'_l}$ follow from Eq.(\ref{IntL4}) of Sect.\ref{Main} on account of the property (\ref{DiffRule}) of $\calE_{\nu, \mu}(z)$, $\partial_z\calE_{\nu,\mu}(z)=\calE_{\nu,\mu+1}(z)$. Here $\Delta(x,x')$ is the dedensitized Van-Vleck determinant (\ref{Van_Vleck}) and we took into account that
\begin{equation}
\det(-\bar\sigma^a_{b'}) = \frac{g^{1/2}(x')}{g^{1/2}(x)}\, \Delta(x,x').
\end{equation}

The new tensor coefficients
\begin{align}
S_{r,l}=S_{r,l,\,a'_1\ldots a'_l}(x,x'),\quad r\geq 0, \quad l = r,\ldots,2r,
\end{align}
are defined by the following simple rules. They are fully symmetric covariant tensors of rank $l$, which consist of $2r-l$ factors $\sigma_{a'}$ and $l-r$ factors of the form $\gamma_{a'b'} = \bar\sigma_{a'}^c g_{cd} \bar\sigma^d_{b'}$, with the combinatorial coefficients equal to the number of different terms in the symmetrization over $l$ indices. For example,
\begin{equation} \label{Stensors}
\begin{aligned}
S_{1,1} &= \sigma_{a'}, & S_{1,2} &= \gamma_{a'b'}, &\!\!\!\!\!\!\!\!\!\!\!\!\!\!\!S_{2,2}= \sigma_{a'} \sigma_{b'},& \\
S_{2,3} &= 3 \gamma_{(a'b'} \sigma_{c')}, & S_{2,4} &= 3 \gamma_{(a'b'} \gamma_{c'd')}&&\\
S_{3,3} &= \sigma_{a'}\sigma_{b'} \sigma_{c'}, \!\!\!& S_{3,4} &= 6 \gamma_{(a'b'} \sigma_{c'} \sigma_{d')}, \!\!\!&&
\end{aligned}
\end{equation}
and so on. They have important property---in the coincidence limit, $\sigma^{a'}=0$, they are nonvanishing only for $l=2r$.

Thus, combining \eqref{bbKRazl} with \eqref{IntL4} we have
\begin{multline} \label{bbKRazl2}
\hat K(\tau| x,x') = \frac{\Delta^{-1}(x,x')\,g^{1/2}(x')}{(4\pi\tau^{1/M})^{d/2}} \sum\limits_{n=0}^\infty \tau^n \sum\limits_{0\le l\le L_n} \\
\times \sum\limits_{r \ge \frac{l}{2}}^l  \calE_{M, \frac{d}{2}+r}\left(-\frac{\sigma}{2\tau^{1/M}}\right) \frac{S_{r,l} * \hat T_{k,l}}{\big(-2\tau^{1/M}\big)^r} \,\hat\calI(x,x') \\
= \frac{\Delta^{-1}(x,x')}{(4\pi\tau^{1/M})^{d/2}} \sum\limits_{n=0}^\infty \!\!\sum\limits_{\;\;\;0\le r\le L_n} \tau^{n-\frac{r}{M}}
\calE_{M, \frac{d}{2}+r}\left(-\frac{\sigma}{2\tau^{1/M}}\right)\\
\times\sum\limits_{l=r}^{\lfloor R_{r,n}\rfloor} \frac{S_{r,l} * \hat T_{n,l}}{(-2)^r}\, \hat\calI(x,x')\,g^{1/2}(x'),
\end{multline}
where we interchanged the order of summation over $l$ and $r$ and introduced the upper limit of summation over $l$---the integer part of
\begin{equation}
R_{r,n} = \min\{2r, L_n(2M)\}.  \label{Rrn}
\end{equation}
Then we replace the summation indices, $(n,r)\to(m,n)$, $m=Mn-r$, under which their summation ranges change like
\begin{equation}
\sum\limits_{n=0}^\infty \sum\limits_{\;\;0\le r\le L_n}
 \!\sum\limits_{l=r}^{\lfloor R_{r,n}\rfloor}\to \sum\limits_{m=-\infty}^\infty \sum\limits_{\;\;n\ge N_m}\!\! \sum\limits_{\;\;\;l=Mn-m}^{\lfloor L_{m,n}\rfloor},
\end{equation}
with $N_m(M)$ given by Eq.(\ref{N_m}) and $L_{m,n}=R_{Mn-m,n}$. After making all these replacements, we get the final expansion (\ref{MainExpansion})-(\ref{Rnm}), where (\ref{Rnm}) follows from (\ref{Rrn}) at $r=Mn-m$ in view of the expression (\ref{Ln}) for $L_n(2M)$.

This is the main result of the paper. For clarity we repeat the expression for this expansion again
\begin{multline} \label{MainExpansion1}
\hat K(\tau| x,x') =  \frac{\Delta^{-1}(x,x')}{(4\pi\tau^{1/M})^{d/2}}\,g^{1/2}(x')
\sum\limits_{m=-\infty}^\infty \tau^{m/M} \\
\times \sum\limits_{\,n \ge N_m}^\infty \calE_{M,\frac{d}2+M n-m}\left(-\frac{\sigma}{2\tau^{1/M}}\right)\, \hat b_{m,n}(x,x'),
\end{multline}
which contains a double series with what we call \emph{the generalized HaMiDeW coefficients} $\hat b_{m,n}(x,x')$ defined by finite sums of Eqs. (\ref{bmn})-(\ref{Rnm}). For a particular subset of indices $m = Mn$ these coefficients, as it follows from (\ref{UltraMarginal}), are universally expressed as
\begin{equation}
\hat b_{Mn,n}(x,x') = \hat T_{n,0}(\nabla)\hat\calI
= \frac{1}{n!}(-\hat F)^n \hat\calI.      \label{special_b}
\end{equation}

The main difference of the expansion (\ref{MainExpansion}) from the Schwinger-DeWitt expansion (\ref{HeatKernel0}) is the presence of infinitely many negative powers of $\tau$ and the absence of one overall exponential function multiplying the proper time series. Instead, every single power $\tau^{m/M}$ and every single generalized HaMiDeW-coefficient $\hat b_{m,n}$ are multiplied by their own generalized exponential function explicitly depending on both indices $m$ and $n$ of the double infinite series. In contrast to anticipations of \cite{Wach2}, the use of generalized exponential functions does not fully perform resummation of all negative powers of $\tau$.

\subsection{Heat kernel diagonal}

Physically interesting and most commonly considered in mathematics is the coincidence limit of the heat kernel. It involves a particular value of the generalized exponential function $\calE_{M,\frac{d}2+M n-m}(0)$ (see \eqref{calE0}) and the coincidence limits $[\,\hat b_{m,n}\,]$. Remarkably, for a given $m$ the latter are vanishing outside of a limited range of $n$,
\begin{equation} \label{inessential}
[\,\hat b_{m,n}\,]= 0,\quad n>4m,
\end{equation}
and, as a result, they are vanishing for all negative $m$, because the summation index $n$ in (\ref{MainExpansion}) is always positive. This property follows from the remark after the formula \eqref{Stensors} that $[\,S_{l,p}\,]\neq 0$ only when $l=2p$, so that the nonvanishing contribution in the sum (\ref{bmn}) over $l$ is possible only when $l=2(Mn-m)\leq L_n(2M)=(2M-1/2)n$.

Thus, the heat kernel coincidence limit has only a finite number of terms with negative powers of $\tau$ and reads
\begin{equation} \label{DiagRazl}
\hat K(\tau| x,x) = g^{1/2}(x)\sum\limits_{m=0}^\infty \tau^\frac{m-d/2}{M} \hat E_{2m}(x),
\end{equation}
where $\hat E_{2m}(x)$ represent the generalization of Seeley--Gilkey coefficients \cite{Seeley, Gilkey1975, Gilkey1979} (usually given under the matrix trace over vector bundle indices and integrated over the spacetime, which often leads to the loss of total derivative terms),
\begin{align} \label{EmCoeff}
&\hat E_{2m}(x) =  \frac{1}{(4\pi)^{d/2}} \sum\limits_{n\geq\frac{m}{M}}^{4m} \frac{\Gamma\left(\frac{d/2-m}{M}+n\right)}{M\Gamma(\frac{d}2-m+M n)} \,[\,\hat b_{m,n}\,],\\
&{\rm dim}\,\hat E_{2m}(x)=2m.
\end{align}
As we see, the coincidence limit represents essential truncation of the expansion (\ref{MainExpansion}), because the coefficient of every power of $\tau$ is given by a finite number of terms.

\begin{figure}
\includegraphics{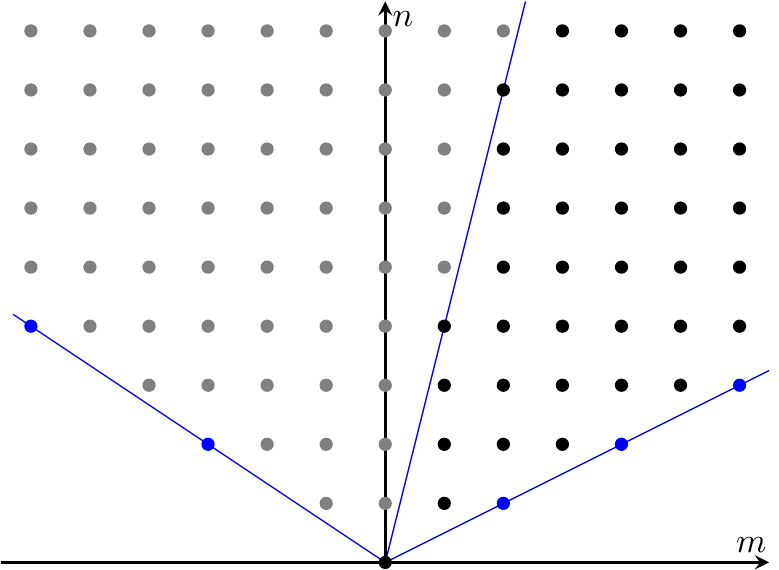}
\caption{\label{FigB} Coefficients $\hat b_{m,n}$ for the case $M=2$. The blue lines at the left and right are given by the expressions (\ref{N_m}) for $N_m(2)$. Black dots show essential coefficients, and gray dots show marginal ones, they are distinguished by the blue line, determined by the condition \eqref{inessential}}.
\end{figure}

It should be emphasized that (\ref{inessential}) is not the only range where the generalized HaMiDeW-coefficients or their coincidence limits are vanishing. As will be shown below, some of the coefficients $\hat b_{m,n}(x,x')$ are identically vanishing for all $x$ and $x'$ due to the properties of $\sigma(x,x')$ and $\hat\calI(x,x')$. These coefficients in the expansion (\ref{MainExpansion}) will be called \emph{spurious}. Non-spurious coefficients, in their turn, can be nonvanishing in the coincidence limit $x=x'$---in this case we will call them \emph{essential}. Other non-spurious coefficients vanishing at $x'=x$, $[\hat a_{m,n}] = 0$,  we will call \emph{marginal}. Thus, in the range of indices  \eqref{inessential} all the coefficients are obviously marginal. Essential and marginal coefficients for the case $M=2$ are shown in Fig. \ref{FigB}.

The coefficients with $m<0$ at separate points $x$ and $x'$  are not automatically excluded  in the expansion (\ref{MainExpansion}) even in the case of the Laplace type operator $M=1$, which seems to contradict the standard Schwinger-DeWitt expansion (\ref{HeatKernel0}). This means that apparently for the $M=1$ case all the coefficients are spurious. This will be explicitly checked below for several low order coefficients and proven for a wide class of minimal higher-derivative operators given by an arbitrary power of the minimal second-order one. Otherwise, the expansion (\ref{MainExpansion}) perfectly passes the consistency test with the Schwinger-DeWitt expansion (\ref{HeatKernel0}) because all of its various generalized exponential kernels for $M=1$ degenerate to a single function $\exp(-\sigma/2\tau)$ in view of the property (\ref{HeatKernel}) of $\calE_{\nu, \mu}(z)$, $\calE_{1,\mu}(z)=e^z$.

\subsection{Causal differential operators}

For the class of {\em causal} operators the coefficient matrix of its highest-derivative term is built of the combination of spacetime metric and Kronecker delta symbols.\footnote{Causality connotation of these operators is associated with the fact that matrix determinant of their symbol reads as a power of the momentum squared, ${\rm det}\,\hat F_N(p)\propto (g^{ab}(x)\,p_ap_b)^{N\tr\hat 1/2}$, so that the characteristic surface of the hyperbolic wave equation in spacetime with the Lorentzian signature, which is defined by the equation $\det\hat F_N(p)=0$, corresponds to the light cone in the metric $g_{ab}$ \cite{Barvinsky85}.} Their principal symbol matrix can be diagonalized in terms of projectors $\hat\varPi_J=\hat\varPi_J(n)$ on different field polarizations, constructed in terms of the unit vector $n_a=p_a/p$, $p=\sqrt{g^{ab}(x)\,p_ap_b}$, and satisfying the orthogonality relations $\hat\varPi_J\,\hat\varPi_K=\delta_{JK}\hat\varPi_J$,
\begin{equation}
\hat F_N(p) = p^N\sum\limits_J f_J \hat\varPi_J.
\end{equation}
Therefore the matrix valued exponential function in (\ref{IntL3}) and (\ref{D}) reduces to the sum of c-number ones with the corresponding purely numerical (independent of $p$) eigenvalues $f_J$,
\begin{equation}
\exp\left(-\tau\hat F_N(p)\right) = \sum\limits_J \hat\varPi_J\,\exp\left(-\tau f_J p^N\right),
\end{equation}
and, in particular, allows one to perform integration in (\ref{D}) in a closed form. At least for even $N=2M$ this diagonalization procedure reduces the calculations to the case of a minimal differential operators of arbitrary even order, for which the expansion formalism can be further developed without restrictions.

\section{Comparison with Schwinger-DeWitt technique} \label{Examples}
In this section, we demonstrate our calculations of the coefficients $\hat b_{m,n}(x,x')$ and their coincidence limits $[\hat b_{m,n}]$ for a minimal operator of the second order,
\begin{equation}
\hat F(\nabla) = -\Box\,\hat 1 + \hat P_1^a\nabla_a + \hat P_0.
\end{equation}
Redefining the connection, $\nabla_a \mapsto \nabla_a + \frac{1}{2} \hat P_{1a}$, allows one to nullify the first-order term, and adding for brevity of the formalism the term $\hat 1R/6$ finally brings $\hat F(\nabla)$ to the form adopted in \cite{Barvinsky85}\footnote{This notation differs from \cite{Barvinsky85} by an overall sign, providing its positivity under Dirichlet boundary conditions at infinity.}
\begin{equation}
\hat F(\nabla) = -\Box\,\hat 1 + \hat P + \frac{\hat 1}6\,R.
\end{equation}

In this case all generalized exponential functions $\calE_{M,d/2-n}\left(-\sigma/2\tau^{1/M}\right)$ degenerate into the usual exponential function $\exp\left(-\sigma/2\tau\right)$ which can be factored out to give the following answer
\begin{multline} \label{LapOpAsympt}
\hat K(\tau| x,x') = \frac{\Delta^{-1}(x,x')\,g^{1/2}(x')}{(4\pi\tau)^{d/2}}\, \exp\left(-\frac{\sigma}{2\tau}\right) \\
\times \sum\limits_{m=-\infty}^\infty \tau^m \hat b_m(x,x'),
\end{multline}
where each of the coefficients $\hat b_m(F| x,x')$ is given by the sum of an infinite number of coefficients $\hat b_{m,n}(F| x,x')$
\begin{equation} \label{LaplaceBcoeficients}
\hat b_m(x,x') = \sum\limits_{n\ge N_m}^\infty \hat b_{m,n}(x,x'),
\end{equation}
where
\begin{equation}
N_m = \begin{cases}
m, & m>0, \\
-2m, & m<0.
\end{cases}
\end{equation}

First of all, using the formulas \eqref{ExpF1} and \eqref{Commut}, we find $\lb\hat F\rb_{m,n}$:
\begin{align}
&\lb\hat F\rb_1^{a'} = -2\sigma^{a'b}\nabla_b - \sigma^{a'b}{}_b, \\
&\lb\hat F\rb_2^{a'_1a'_2} = \sigma^{a'_1}{}_b \sigma^{a'_2b}, \\
&\lb\hat F\rb_{1,1}^{a'_1a'_2a'_3} = 2i\sigma^{a'_1b_1}\nabla_{b_1}(\sigma^{a'_2}_{b_2}\sigma^{a'_3b_2}), \\
&\lb\hat F\rb_{0,2}^{a'_1a'_2a'_3a'_4} = -\nabla_{b_3}(\sigma^{a'_1}_{b_1} \sigma^{a'_2b_1}) \nabla^{b_3}(\sigma^{a'_3}_{b_2} \sigma^{a'_4b_2}),\\
&\lb\hat F\rb_{0,1}^{a'_1a'_2} = 2\nabla^{b_2}(\sigma^{a'_1}_{b_1}\sigma^{a'_2b_1})\nabla_{b_2}
+\nabla_{b_2}\nabla^{b_2}(\sigma^{a'_1}_{b_1}\sigma^{a'_2b_1}).
\end{align}

Let us find the expressions we need for $\hat T_{k,l}$ by the formula \eqref{FrakTRecRel} (the coefficients with $l=0,1$ are given by the formula \eqref{ProdOpBrak}, and the coefficients of the form $\hat T_{2k,3k}$ are given by the formula \eqref{UltraMarginal})
\begin{align}
&\hat T_{2,2} = \frac{1}{2}\left(\lb\hat F\rb_1^2 - \lb\hat F\rb_{0,1}\right),\\
&\hat T_{3,2} = -\frac{1}{3!}\left(\hat F \lb\hat F\rb_1^2 + \lb\hat F\rb_1\lb\hat F^2\rb_1- \hat F\lb\hat F\rb_{0,1}\right. \nonumber\\
&\qquad\qquad\left. - 2\lb\hat F\rb_{0,1}\hat F\right),\\
&\hat T_{3,3} = -\frac{1}{3!}\left(\lb\hat F\rb_1^3 - \lb\hat F\rb_1\lb\hat F\rb_{0,1} - 2\lb\hat F\rb_{0,1}\lb\hat F\rb_1 \right. \nonumber\\
&\qquad\qquad\left.- \hat F\lb\hat F\rb_{1,1} - 2\lb\hat F\rb_{1,1}\hat F\right),\\
&\hat T_{3,4} = \frac{1}{3!}\left(\lb\hat F\rb_1 \lb\hat F\rb_{1,1} + 2\lb\hat F\rb_{1,1} \lb\hat F\rb_1- 2\lb\hat F\rb_{0,2}\right),
\end{align}

\paragraph{Negative powers.} This expansion differs from the standard DeWitt expansion by the presence of negative powers of $\tau$ ($m<0$). This visible discrepancy is removed by the observation that the coefficients $\hat b_{m,n}$ for $m<0$ are in fact spurious, i.e. they vanish identically not only in the coincidence limit, but also with the separated points $x\neq x'$. Although we could not prove this property in general, we verified it by direct calculations for the first four coefficients of this form (i.e. for $\hat b_{-1,2}$, $\hat b_{-1,3}$, $\hat b_{-1,4}$ and $\hat b_{-2,4}$). It turns out that by virtue of the relation $\sigma_{a'bc}\sigma^b\sigma^c \equiv 0$ \eqref{SigmaSymm} and other more complicated relations of the same type, these four first coefficients really vanish identically. Here are the explicit expressions
\begin{align}
&\hat b_{-1,2} = -\frac{1}{4} \sigma_{a'bc}\sigma^{a'}\sigma^b\sigma^c \hat \calI \equiv 0,\\
&\hat b_{-1,3} = \frac{1}{12} \sigma^{a'_1}\sigma^{a'_2}\sigma^{b_1}\sigma^{b_2} (\sigma_{a'_1}{}^c{}_{b_1}\sigma_{a'_2cb_2} \nonumber\\
&\qquad\quad+ \sigma_{a'_1b_1}{}^c\sigma_{a'_2cb_2} + \sigma_{a'_1b_1}{}^c\sigma_{a'_2b_2c})\, \hat \calI \equiv 0,\\
&\hat b_{-2,4} = \frac{1}{32} \left(\sigma_{a'bc}\sigma^{a'}\sigma^b\sigma^c\right)^2 \hat \calI \equiv 0.
\end{align}
We will not give an expression for the coefficient $\hat b_{-1,4}$ due to its complexity, but only indicate that it contains convolutions of the form
\begin{equation}
\sigma^{a'_1}\sigma^{a'_2}\sigma^{a'_3}\sigma^{b_1}\sigma^{b_2} \sigma_{a'_1b_1c_1}\sigma_{a'_2b_2c_2} \sigma_{a'_3}{}^{c_1c_2}.
\end{equation}
All terms of this kind are also identically zero.

\paragraph{Positive powers.} We now turn to the discussion of positive powers. Another discrepancy between our expansion \eqref{LaplaceBcoeficients} and the standard expansion \eqref{HeatKernel0} is that instead of the standard prefactor $\Delta^{1/2}(x,x')$ we have the inverse determinant $\Delta^{-1}(x,x')$. Therefore, the following relation should hold between the original Schwinger-DeWitt coefficients and the generalized ones
\begin{multline} \label{Prefactor}
\hat a_m(x,x') = \Delta^{-3/2}(x,x')\, \hat b_m(x,x')\\
=\Delta^{-3/2}(x,x')\!\sum\limits_{n\ge N_m}^\infty
\hat b_{m,n}(x,x'), \quad m\ge0.
\end{multline}

Note that if we are only interested in the coincidence limits then, as it follows from \eqref{EmCoeff}, only the first $3m+1$ terms (up to $k=4m$) from the whole infinite series \eqref{LaplaceBcoeficients} will contribute to $[\hat b_m]$. (Actually, as we will see later, even less due to symmetry reasons.)

We compare now the results obtained by our method with the well-known results for the coincidence limits of the first HaMiDeW coefficients and their derivatives (they can be found in \cite{DeWitt} or in \cite{Barvinsky85}),
\begin{align}
&[\hat a_0] = \hat 1, \quad [\nabla_a\hat a_0] = 0, \quad [\nabla_a\nabla_b \hat a_0] = \frac{1}{2}\hat\calR_{ab}, \label{A0Coeff} \\
&[\hat a_1] = -\hat P, \quad [\nabla_a\hat a_1] = -\frac{1}{2}\nabla_a\hat P - \frac{1}{6}\nabla^b\hat\calR_{ba}, \label{A1Coeff}\\
&[\hat a_2] = \frac{1}{180}\left(R_{abcd}R^{abcd} - R_{ab}R^{ab} + \Box R\right)\hat 1 \nonumber\\
&\qquad\quad+ \frac{1}{2}\hat P^2 + \frac{1}{12}\hat\calR_{ab}\hat\calR^{ab} -\frac{1}{6}\Box\hat P. \label{A2Coeff}
\end{align}
In order to compare the coincidence limits for their derivatives, one would need the coincidence limits for the Van Vleck prefactor $\Delta^{-3/2}$ in \eqref{Prefactor},
\begin{equation} \label{DeltaLimits}
[\Delta] = 1, \quad [\nabla_a\Delta] = 0, \quad [\nabla_a\nabla_b\Delta] = \frac{1}{3}R_{ab}.
\end{equation}

Let us start with the coefficients with $m=0$. Two-point expressions for several lowest $\hat b_{0,n}(x,x')$ are
\begin{eqnarray}
&&\hat b_{0,0} = \hat\calI,\quad \hat b_{0,1} = -\frac{1}{2}\sigma_{a'} \sigma^{a'b}{}_b \,\hat\calI, \\
&&\hat b_{0,2} = \frac{1}{2} \sigma^{b'} \sigma^c \left(\sigma_{b'ac} + \sigma_{b'ca}\right) \hat\calI_a
+ \frac{1}{8} \left(\sigma^{a'} \big(4\sigma_{a'b}{}^b\right.\nonumber \\
&&\qquad+ \sigma^{b'} (2\sigma_{a'}{}^{cd}\sigma_{b'cd} + \sigma_{a'c}{}^c \sigma_{b'd}{}^d)
+ 2\sigma^b (\sigma_{a'bc}{}^c + \sigma_{a'c}{}^c{}_b)\big) \nonumber\\
&&\qquad\left. + 4\sigma^a \bar\sigma_{cb'} \big(\sigma^{b'}{}_a{}^c + \sigma^{b'c}{}_a\big)\right) \hat\calI.
\end{eqnarray}
Their coincidence limits and coincidence limits of their lowest order derivatives immediately follow as
\begin{align}
&[\hat b_{0,0}] = \hat 1, \;\; [\nabla_a \hat b_{0,0}] = 0, \;\; [\nabla_a\nabla_b \hat b_{0,0}] = \frac{1}{2}\hat\calR_{ab}, \\
&[\hat b_{0,1}] = 0, \;\;  [\nabla_a \hat b_{0,1}] = 0, \;\;  [\nabla_a\nabla_b \hat b_{0,1}] = \frac{2}{3}R_{ab}\hat 1 ,\\
&[\hat b_{0,2}] = 0, \;\;  [\nabla_a \hat b_{0,2}] = 0, \;\;  [\nabla_a\nabla_b \hat b_{0,2}] = -\frac{\hat 1}{6}\, R_{ab}.
\end{align}
We do not give explicit expressions for the coefficients $\hat b_{0,3}$ and $\hat b_{0,4}$ due to their complexity (for example, the expression for the latter contains about 750 terms), however, for their coincidence limits, the calculation gives:
\begin{align}
&[\hat b_{0,3}] = [\hat b_{0,4}] = 0, \\
&[\nabla_a\hat b_{0,3}] = [\nabla_a\hat b_{0,4}] = 0, \\
&[\nabla_a\nabla_b\hat b_{0,3}] = [\nabla_a\nabla_b\hat b_{0,4}] = 0.
\end{align}

Collecting these results we obtain for $m=0$
\begin{equation}
[\hat b_0] = \hat 1, \quad [\nabla_a \hat b_0] = 0, \quad [\nabla_a\nabla_b \hat b_0] = \frac{1}{2}\big(\hat\calR_{ab} + \hat 1 R_{ab}\big).
\end{equation}
Using these relations in Eq.\eqref{Prefactor} for $\hat a_m(x,x')$ in terms of $\hat b_{m,n}(x,x')$ we recover on account of  \eqref{DeltaLimits} the expressions \eqref{A0Coeff} well known from the solution of DeWitt recursion relations.

Similarly, the results for the first two $\hat b_{1,n}(x,x')$
\begin{align}
&\hat b_{1,1} = -\hat F(\nabla) \hat\calI, \\
&\hat b_{1,2} = -\frac{1}{2}\sigma_a\big(\hat\calI^{ab}{}_b + \hat\calI_b{}^{ba}\big) - \frac{1}{2}\big(2 + \sigma^{a'}\sigma_{a'b}{}^b\big)\,\hat\calI_c^c \nonumber\\
&- \sigma^{a'}\sigma_{a'bc}\,\hat\calI^{bc} - \frac{1}{2}\Big(\sigma^{a'}(\sigma_{a'bc}{}^c + \sigma_{a'c}{}^c{}_b) + 2\sigma^{a'c}{}_c\bar\sigma_{ba'} \nonumber\\
&+ 2(\sigma_{a'bc} + \sigma_{a'cb})\bar\sigma^{ca'}\Big)\,\hat\calI^b - \frac{1}{4}\Big(2(\sigma^{a'bc}{}_c + \sigma^{a'c}{}_c{}^b)\bar\sigma_{ba'} \nonumber\\
&+ \sigma^{a'}\sigma_{a'b}{}^{bc}{}_c + (\sigma^{a'c}{}_c\sigma_{b'd}{}^d + 2\sigma^{a'cd}\sigma_{b'cd})\bar\sigma_{ea'}\bar\sigma^{eb'}\Big)\,\hat\calI \nonumber\\
&+ \frac{1}{2}\Big(\sigma^{a'}\sigma_{a'b}{}^b\big(\hat P
+ \frac{\hat 1}{6}R\big) + \sigma^a\big(\nabla_a\hat P
+ \frac{\hat 1}{6}\nabla_aR\big)\Big)\,\hat\calI,
\end{align}
lead to the coincidence limits
\begin{align}
&[\hat b_{1,1}] = -\hat P - \frac{\hat 1}{6}R, \\
&[\nabla_a \hat b_{1,1}] = -\nabla_a\hat P - \frac{\hat 1}{6}\nabla_aR - \frac{1}{3}\nabla^b\hat\calR_{ab},\\
&[\hat b_{1,2}] = \frac{\hat 1}{6}\,R, \\
&[\nabla_a\hat b_{1,2}] = \frac{1}{2}\nabla_a\hat P
+ \frac{\hat 1}{3}\nabla_aR + \frac{1}{6}\nabla^b\hat\calR_{ab}.
\end{align}
We do not present the expressions for $\hat b_{1,3}$ and $\hat b_{1,4}$ (for example, the expression for the latter contains about 3500 terms), but their coincidence limits are very concise,
\begin{align}
&[\hat b_{1,3}] = [\hat b_{1,4}] = 0, \\
&[\nabla_a\hat b_{1,3}] = - \frac{\hat 1}{6}\,\nabla_aR,
\quad [\nabla_a\hat b_{1,4}] = 0.
\end{align}

Collecting these results in (\ref{LaplaceBcoeficients}), we obtain for $m=1$
\begin{equation}
[\hat b_1] = - \hat P, \quad [\nabla_a \hat b_1] = -\frac{1}{2}\nabla_a\hat P + \frac{1}{6}\nabla^b\hat\calR_{ab},
\end{equation}
which again confirm the relations \eqref{A1Coeff} for the HaMiDeW coefficient $\hat a_1(x,x')$.

Finally, consider the results for the coincidence limits of the first coefficients $\hat b_{2,n}(x,x')$ needed for the verification of the known result for $\hat a_2(x,x)$,
\begin{align}
&[\hat b_{2,2}] = \frac{1}{2}\Big(\hat P + \frac{\hat 1}{6}R\Big)^2 =\frac{1}{2}\Box\Big(\hat P + \frac{\hat 1}{6}R\Big) \nonumber\\
&\qquad\quad+ \frac{1}{4}\hat\calR_{ab}\hat\calR^{ab},\\
&[\hat b_{2,3}] = \frac{1}{540}\,(12R_{abcd}R^{abcd} - 12R_{ab}R^{ab} - 15R^2 \nonumber\\
&\quad\quad- 102\Box R)\hat 1 - \frac{1}{6}\,(\hat\calR_{ab}\hat\calR^{ab} + R\hat P + 2\Box\hat P ),\\
&[\hat b_{2,4}] = \frac{1}{360}\,(-6R_{abcd}R^{abcd} + 6R_{ab}R^{ab} + 5R^2\nonumber\\
&\qquad\quad+ 36\Box R), \\
&[\hat b_{2,5}]=[\hat b_{2,6}]=[\hat b_{2,7}]=[\hat b_{2,8}]=0.
\end{align}
The sum of these expressions exactly generates the well-known expression \eqref{A2Coeff} which plays a very  important role because it exhausts one-loop divergences of a generic 4-dimensional field theory.

Our checks give sufficient evidence that the application of the generalized Fourier method, in all parts, yields the results coinciding with those of the standard DeWitt technique. There are, however, differences between these methods. In the Schwinger-DeWitt technique to calculate the coincidence limit $[\hat a_m]$ one needs to know the coincidence limits of all lower $m$ coefficients and a number their derivatives. On the contrary, in our method, the recurrent procedure is carried out not for the coefficients $\hat b_{m,n}$ themselves, but for the operators $\hat T_{n,l}(\nabla)$, and the coefficients are obtained independently of each other in the form of contractions of these operators with the tensors $S_{r,l}$ by Eq. \eqref{bmn}. Moreover, this gives closed expressions for the coefficients $\hat b_{m,n}(x,x')$ at separate points $x\ne x'$ and via Eq.(\ref{Prefactor}) yields entirely new representation for HaMiDeW coefficients.

\section{Fourth order minimal operator} \label{4thOrder}

Here we consider a generic minimal fourth order operator,
\begin{equation} \label{Gen4Order}
\hat F(\nabla) = \Box^2 + \hat\varOmega^{abc}\nabla_a\nabla_b\nabla_c+
\hat D^{ab}\nabla_a\nabla_b+H^a\nabla_a+\hat P,
\end{equation}
where $\hat\varOmega^{abc} = \hat\varOmega^{(abc)}$ and $\hat D^{ab} = \hat D^{(ab)}$. Relevant computations and especially the representation of the final results become excessively involved in this case if the coefficient $\hat\varOmega^{abc}$, ${\rm dim}\,\hat\varOmega^{abc}=1$, is nonzero. Therefore we present here the results for the case of $\hat\varOmega^{abc}=0$ relegating to Appendix \ref{Omega} the generalization to nonzero value of this coefficient. Again, the chain of differential operators $\hat T_{n,l}(\nabla)$ and generalized Schwinger-DeWitt coefficients $\hat b_{m,n}(x,x')$ are so complicated, that we will skip initial calculational steps and go directly to the results for several lowest order coincidence limits. For the coefficients of dimensionality two they are rather simple ($\hat D = \hat D_{a}^a$),
\begin{align}
[\hat b_{1,1}] &= \frac{1}{2} \hat D - \frac{\hat 1}{3}R, \\
[\hat b_{1,2}] &= \frac{\hat 1}{6}(d+2)(d+4)R,\\
[\hat b_{1,3}] &= [\hat b_{1,4}]=0,
\end{align}
whereas for dimensionality four they involve all operator coefficients and read
\begin{align}
[\hat b_{2,1}] =&-\hat P-\frac{1}{2}\hat\calR_{ab}\hat\calR^{ab},\nonumber\\
[\hat b_{2,2}]=& \frac{\hat 1}{15} (2d+7) \Big(R_{ab}^2 - R_{abcd}^2- 6\Box R \Big)+ \frac{\hat 1}{6} R^2 \nonumber\\
&+ \frac{1}{4}(d+2)(d+4) \hat\calR_{ab}^2+ \frac{1}{8} \hat D^2 + \frac{1}{4} \hat D_{ab} \hat D^{ab} \nonumber\\
&- \frac{1}{3}(d+3) \hat D^{ab} R_{ab} -  \frac{1}{6} \hat D R + \frac{1}{2} (d+2) \nabla_a\hat H^a\nonumber\\
&+ \nabla_a\nabla_b \hat D^{ab} + \frac{1}{4} (d+4) \Box \hat D, \\
[\hat b_{2,3}] =& - (d+4)(d+6) \Big( \frac{d+11}{45} \big(R_{ab}^2 - R_{abcd}^2\nonumber\\
&- 6\Box R \big)\,\hat 1 + \frac{\hat 1}{6} R^2  + \frac{1}{6}(d+2) \hat\calR_{ab}^2- \frac{1}{6} \hat D^{ab} R_{ab} \nonumber\\
&- \frac{1}{12} \hat D R + \frac{1}{3} \nabla_a\nabla_b \hat D^{ab} + \frac{1}{6} \Box\hat D \Big), \\
[\hat b_{2,4}] = &(d+4) (d+6) (d+8) (d+10)\nonumber\\
&\times\Big(\frac{\hat 1}{60} \big(R_{ab}^2 - R_{abcd}^2 - 6\Box R\big) + \frac{\hat 1}{72} R^2 \Big),
\end{align}
\begin{equation}
[\hat b_{2,5}] = [\hat b_{2,6}]=[\hat b_{2,7}]=[\hat b_{2,8}].
\end{equation}

These results show that the coincidence limits $[\hat b_{m,n}]$ for a given $m$ start vanishing essentially earlier than at $n>4m$. This happens due to the properties of symmetrized covariant derivatives of Synge world function and, apparently, starts at $n>2m$, though we do not yet have the proof of this property in higher orders.

Thus we have the Seeley--Gilkey coefficients (\ref{EmCoeff}) of the heat kernel diagonal of the minimal fourth order operator. The dimensionality two coefficient is contributed by nonvanishing $[\hat b_{1,1}]$ and $[\hat b_{1,2}]$ and reads
\begin{align} \label{E2Coeff}
\hat E_2(x) =\frac{1}{(4\pi)^{d/2}} \frac{\Gamma\left(\frac{d/2-1}{2}\right)}{2\Gamma(\frac{d}2-1)} \bigg\{\frac{1}{2d}\hat D + \frac{\hat 1}{6}R\bigg\}.
\end{align}
The dimensionality four coefficient includes nonvanishing contributions of the above $[\hat b_{2,1}]$,...,$[\hat b_{2,4}]$ coincidence limits and equals
\begin{align}
\hat E_4(x) &= \frac{1}{(4\pi)^{d/2}} \frac{\Gamma\left(\frac{d}{4}\right)}{4\Gamma\left(\frac{d}{2}\right)}
\Bigg\{(d-2)\Big(\frac{\hat 1}{90}R^2_{abcd}-\frac{\hat 1}{90}R^2_{ab}
\nonumber\\
&+\frac{\hat 1}{36} R^2 + \frac{1}{6} \hat\calR^2_{ab}
+ \frac{\hat 1}{15}\,\Box R\,\Big)
- \frac{1}{3} \hat D^{ab} R_{ab} + \frac{1}{6} \hat D R \nonumber\\
&+ \frac{1}{d+2} \Big(\frac{1}{2}\hat D_{ab} \hat D^{ab} + \frac{1}{4}\hat D^2 - \frac{2}{3}(d+1) \nabla_a\nabla_b \hat D^{ab} \nonumber\\
&+ \frac{1}{6}(d+4) \Box \hat D\Big) - 2\hat P + \nabla_a \hat H^a\Bigg\}. \label{E4Coeff}
\end{align}

Both expressions fully correspond to the results of \cite{Barvinsky85,Gusynin1990}. The paper \cite{Barvinsky85} contains also the contribution of nonzero $\varOmega^{abc}$, but without the total derivative terms. In view of very lengthy terms with $\varOmega^{abc}$ their complete contribution is given in Appendix \ref{Omega}.

The off-diagonal heat kernel of (\ref{Gen4Order}) is too complicated to be presented here even in the lowest orders of its expansion. Therefore we give here only its part which should be indicative of such new features of higher derivative operators as the origin of negative powers of the proper time. This is the marginal (that is vanishing on the heat kernel diagonal, but non-zero outside of it) coefficient $\hat b_{-1,1}(x,x')$ of the negative power of $\tau$.

Using (\ref{bmn}) and (\ref{Rnm}) we have the integer part of $\lfloor L_{-1,1}\rfloor=3$, so that
\begin{equation}
\hat b_{-1,1}(x,x')=
-\frac18\,S_{3,3}*T_{1,3}(\nabla)\,\calI(x,x'),   \label{b-11}
\end{equation}
where the tensor $S_{3,3}$ is defined in Eq.(\ref{Stensors}) and the differential operator $T_{1,3}(\nabla)=-\lb F(\nabla)\rb_{3,0}=-\lb F(\nabla)\rb_3$, as it follows from (\ref{FrakTRecRel})--(\ref{InitT}). So it turns out to be
\begin{align}
T_{1,3}^{a'b'c'}(\nabla) &= -4\sigma^{a'b}\sigma^{b'}_b\sigma^{c'c}\nabla_c
- 2\sigma^{a'b}\sigma^{b'}_b \sigma^{c'd}{}_d \nonumber\\
&- 4\sigma^{a'bc}\sigma^{b'}_b\sigma^{c'}_c
- \varOmega^{abc}\sigma^{a'}_a\sigma^{b'}_b\sigma^{c'}_c.
\end{align}
Therefore, in view of the parallel transport equation for $\calI(x,x')$ and the relation $\sigma^{ab'}\sigma_{b'}=\sigma^a$ we have
\begin{equation} \label{NegPow}
\hat b_{-1,1}(x,x') = \frac{1}{8}\,(4\sigma\sigma_{a'}\Box\sigma^{a'} + \hat \varOmega^{abc}\sigma_a\sigma_b\sigma_c)\, \hat\calI(x,x'),
\end{equation}
which is explicitly non-vanishing for $x\neq x'$. This proves that negative powers of the proper time indeed arise in the expansion (\ref{MainExpansion}) for $M>1$.

Calculations of this and the previous sections show that their complexity is much higher than in the recursive procedure of the Schwinger-DeWitt technique. Without using computer symbolic manipulations, that were performed by means of {\em Wolfram Mathematica} and tensor computer algebra packages {\em xAct} and {\em xTras}, a completion of the above checks would be impossible. Obviously the level of complexity grows if one goes to higher derivative operators with $M>1$. In connection with this we would like to point out to an unsolved and somewhat mysterious feature of our formalism, which followed from the above computer simulations throughout various consistency checks of the above type.

Interesting point is that there exist two versions of the generalized Fourier transform. One of them is covariant and corresponds to the delta-function representation (\ref{DeltaInt}), whereas the non-covariant one proceeds with $k_{a'}\sigma^{a'}(x',x)$ replaced with $k_{a}\sigma^{a}(x,x')$. This means that the auxiliary momentum space tangent to spacetime manifold is introduced at the point $x$ rather than at $x'$.  The non-covariant version has the calculational advantage of having in the game only one unprimed type of indices, but it violates the basic commutation equation (\ref{commutation}) by extra non-covariant terms proportional to the covariant derivatives of the diffeomorphisms inert momentum vector, $(\nabla_b k_a)\sigma^{a}(x,x')=-k_c\varGamma_{ba}^c\sigma^a(x,x')$. Paradoxically, naive omission of these terms has led in the computer simulation of the above type to exactly the same coincidence limits, and this happens despite the fact that $\sigma^a(x,x')$ in the expression above gets numerously differentiated and thus is nonzero in this limit. Thus far we do not have explanation for this digitally generated observation, which however might be useful in symbolic computations.

\section{Powers of a~minimal operator \label{PowerMin}}

In this section we will study the direct and inverse Mellin transform in heat kernel theory which allows one to relate the curvature expansions for the minimal operator $\hat F(\nabla)$ and its integer or non-integer power $\hat F^s(\nabla)$. Since we will have the formalism relating heat kernels, the Green function and $\hat b_{m,n}$ coefficients for different operators, we will mark them by additional subscript or superscript.

The Green function of the operator $\hat F^s(\nabla)$ or the $s$-th power of the Green function of $\hat F(\nabla)$ is given by the Mellin transform---the proper time integral of the form
\begin{align} \label{FnGr}
\hat G_{F^s}(x, x')&\equiv \frac{\hat 1}{F^s(\nabla)}
\delta(x,x')\nonumber\\
&= \frac{1}{\Gamma(s)} \int\limits_0^\infty d\tau\, \tau^{s-1} \hat K_F(\tau| x, x').
\end{align}
The inverse Mellin transform allows one to recover the heat kernel of the operator $\hat F(\nabla)$ from $\hat G_{F^s}(x, x')$
\begin{equation} \label{KFFs}
\hat K_F(\tau| x, x') = \frac{1}{2\pi i} \int\limits_{w-i\infty}^{w+i\infty} ds\,\tau^{-s}\, \Gamma(s)\, \hat G_{F^s}(x, x')\,,
\end{equation}
where the integration runs along the contour parallel to imaginary axis in the complex plane of $s$ for sufficiently big positive $w$. We do not discuss the range of $s$ where both transforms are well defined and only assume positive-definiteness of $\hat F$ and the possibility to make needed analytic continuation. Brief derivation of these transforms  is given in Appendix \ref{GeneralTheory}.

The idea of relating heat kernel curvature expansions for $\hat F^s(\nabla)$ and $\hat F(\nabla)$ is based on the sequence of transformations
\begin{align}  \label{diagram1}
\xymatrix{
\hat K_F(\tau| x, x') \ar[r]^{\calM_{\tau t}} & \hat G_{F^t}(x, x') \ar[d]^{F\to F^s}\\
\hat K_{F^s}(\tau| x, x') &\ar[l]^{\calM^{-1}_{t\tau}} \hat G_{F^{st}}(x,x'),
}\end{align}
where ${\calM_{\tau t}}$ denotes the Mellin transform from the function of the proper time $\tau$ to its image as a function of $t$ and ${\calM^{-1}_{t\tau}}$ is the respective inverse Mellin transform.

\subsection{The Green function of the operator $\hat F^t$.}

Using \eqref{FnGr} and integrating term by term the expansion \eqref{MainExpansion} we get
\begin{multline} \label{Fs}
\hat G_{F^t}(x,x') = \Delta^{-1}(x,x')\, g^{1/2}(x') \\
\times \sum\limits_{m=-\infty}^\infty \sum\limits_{\,n \ge N_m}^\infty \bbG^{m,n}(t, \sigma(x,x'))\, \hat b^F_{m,n}(x,x'),
\end{multline}
where $\hat b^F_{m,n}(x,x')$ are the generalized HaMiDeW coefficients for the operator $\hat F$ and their coefficients, which one might call \emph{the basis Green functions}, equal
\begin{multline} \label{OmegaN}
\bbG^{m,n}(t, \sigma) =\frac{1}{(4\pi)^{d/2}} \left(\frac{\sigma}{2}\right)^{-s} \frac{M}{\Gamma(t)}  \\
\times\int\limits_0^\infty dz\, z^{s-1} \calE_{M,d/2+Mn-m}(-z) \\
= \frac{1}{(4\pi)^{d/2}} \frac{\Gamma\left(\frac{d}{2}-m-Mt\right) \Gamma(n+t)}{\Gamma(t)\,\Gamma\big(M(n+t)\big)} \left(\frac{\sigma}{2}\right)^{Mt + m -\frac{d}{2}}.
\end{multline}
In the derivation of this expression we changed the integration variable, $\tau\to z = \sigma/2\tau^{1/M}$ and used the Mellin transform of the generalized exponential function --- Eq.\eqref{MellinCalE} with $s = d/2 - m - Mt$.

\subsection{The heat kernel of the operator $\hat F^s$.}

The inverse Mellin transform (\ref{KFFs}) now gives the result of the last step in the diagram (\ref{diagram1})---the asymptotic heat kernel expansion for the operator $\hat F^s$,
\begin{multline} \label{KFFs0}
\hat K_{F^s}(\tau| x, x') = \frac{1}{2\pi i} \int\limits_{w-i\infty}^{w+i\infty}dt\,\tau^{-t}\, \Gamma(t)\, \hat G_{F^{st}}(x, x').
\end{multline}

Term by term integration of the expansion \eqref{Fs} with $t\to st$ gives new coefficient functions of $\hat b_{m,n}$
\begin{multline} \label{bbKn}
\frac{1}{2\pi i} \int\limits_{w-i\infty}^{w+i\infty} dt\,\tau^{-t}\, \Gamma(t)\, \bbG^{m,n}(st,\sigma)\, \\
= \frac{\tau^{m/sM}}{(4\pi\tau^{1/sM})^{d/2}}\,
\calE_{M, d/2+Mn-m}^{s, d/2-m}\left(-\frac{\sigma}{2\tau^{1/sM}}\right).
\end{multline}
where $\calE_{\beta,b}^{\alpha,a}(-z)$ are the {\em generalized exponential functions of the second order} \eqref{GEF2order}--- some new hyper-geometric type functions. Their properties and the derivation of this equation are briefly presented in Appendices \ref{GEF} and \ref{GeneralTheory}. As a result
\begin{multline} \label{bbKRazlLapl}
\hat K_{F^s}(\tau| x,x') = \frac{\Delta^{-1}(x,x')}{(4\pi\tau^{1/sM})^{d/2}} \, g^{1/2}(x')
\sum\limits_{m=-\infty}^\infty \tau^{m/sM} \\
\times \sum\limits_{\,n \ge N_m}^\infty \calE_{M, d/2+Mn-m}^{s, d/2-m}\left(-\frac{\sigma}{2\tau^{1/sM}}\right)\,\hat b^F_{m,n}(x,x').
\end{multline}

This relation for off-diagonal elements of heat kernels is also very interesting for the coincidence limit $x'=x$ case. In view of the expression \eqref{H0} for $\calE_{\beta,b}^{\alpha,a}(0)$ the second order generalized exponential functions here reduce to the gamma-function factors in Eq.\eqref{EmCoeff} modulo independent of $n$ coefficient $\Gamma(\frac{d/2-m}{sM})/s\Gamma (\frac{d/2-m}M)$, so that the Seeley--Gilkey coefficients of the operator $\hat F^s$,
\begin{equation} \label{CoincidenceLimit}
\hat K_{F^s}(\tau| x,x) = g^{1/2}(x) \sum\limits_{m=0}^\infty  \tau^\frac{m-d/2}{sM} \hat E^{F^s}_m(x),
\end{equation}
express via those of $\hat F$,
\begin{equation} \label{Acoeff}
\hat E^{F^s}_m(x) = \frac{\Gamma\left(\frac{d/2-m}{sM}\right)}{s\,\Gamma \left(\frac{d/2-m}{M}\right)}\, \hat E^F_m(x).
\end{equation}
This is an extension of a well known Fegan--Gilkey relation between functional traces of these two operators \cite{GilkeyFegan} to the case of their diagonal elements---functions of a spacetime point $x$.

This formula can be, in particular, used to check the consistency of Eqs.(\ref{E2Coeff})-(\ref{E4Coeff}) for Seeley--Gilkey coefficients of the fourth order operator. A squared version $\hat F^2(\nabla)$ of a minimal second order operator $\hat F(\nabla)=-\Box+\hat P$ is the fourth order operator (\ref{Gen4Order}) with
\begin{align}
&\hat\varOmega^{abc}_{F^2}=0, \quad \hat D^{ab}_{F^2}=-2g^{ab}\hat P,
\\
&\hat H^a_{F^2}=-2\nabla_a\hat P, \quad \hat P_{F^2}=\hat P^2-(\Box\hat P).
\end{align}
Substituting these expressions into Eqs.(\ref{E2Coeff})-(\ref{E4Coeff}) one easily confirms Fegan-Gilkey relations (\ref{Acoeff}) between $E^{F^2}_{2,4}(x)$ and $E^{F}_{2,4}(x)$.

The above relations hold also for non-integer powers $s$, for which case we do not see the presence of $\log\tau$-terms advocated to be possible in  mathematical literature \cite{GilkeyGrubb, Bar2003}.  The presence of such terms occurring due to merging poles might be an artifact of the method which was carried out only in the coincidence limit $x=x'$ and required using the $\zeta$-functional regularization. In contrast, our method operates with objects at separated points and does not require additional regularization. The difference between the two methods can be clearly visualized by the lack of closure of the following diagram, where $\hat\zeta_{F}(s, x)=\hat G_{F^s}(x,x)$ denotes the regularized zeta-function in the coincidence limit $x'=x$,
\begin{widetext}
\begin{equation}
\xymatrix{
\hat K_F(\tau| x, x') \ar[r]^{\calM_{\tau t}} \ar[d]^{x\to x'} & \hat G_{F^s}(x, x') \ar[r]^{F\to F^s} & \hat G_{F^{st}}(x, x') \ar[r]^{\calM^{-1}_{t\tau}} & \hat K_{F^s}(\tau| x, x') \ar[d]^{x\to x'} \\
\hat K_F(\tau| x, x) \ar[r]^{\calM_{\tau t}} & \hat\zeta_F(t, x) \ar[r]^{F\to F^s} & \hat\zeta_{F^s}(t, x) \ar[r]^{\calM^{-1}} & \hat K_{F^s}(\tau| x, x).
}\end{equation}
\end{widetext}

\subsection{Integer powers of a Laplace type operator}

Thus far we considered generic fractional and even complex powers $s$ of the operator. Now consider the integer power $s=M$ of the minimal second order operator $\hat F(\nabla)$ and note that this time $\hat F(\nabla)$ and $\hat F^M(\nabla)$  are both minimal differential operators. Then for the operator $\hat F^M(\nabla)$ we have two different expansions: one obtained using the general Fourier method \eqref{MainExpansion}, and another obtained from the expansion for the operator $\hat F(\nabla)$ using the Mellin transform \eqref{bbKRazlLapl}. It is natural to compare these expansions. For the sake of simplicity, we will consider the case when $\hat F^M(\nabla)$ is a Laplace type operator (as we showed above in Sect. \ref{Examples}, the expansion which in this case generates the Fourier method is equivalent to the standard DeWitt expansion).

The Fourier method expansion \eqref{MainExpansion} gives
\begin{multline} \label{MainExpansionF^Z}
\hat K_{F^M}(\tau| x,x') =  \frac{\Delta^{-1}(x,x')}{(4\pi\tau^{1/M})^{d/2}}\,g^{1/2}(x') \sum\limits_{m=-\infty}^\infty \tau^{m/M} \\
\times\!\!\!\! \sum\limits_{\,n \ge N_m(M)}^\infty \calE_{M,\frac{d}2+M n-m}\left(-\frac{\sigma}{2\tau^{1/M}}\right)\, \hat b^{F^M}_{m,n}(x,x'),
\end{multline}
while exponentiation using the Mellin transform leads to the formula (which is a special case of \eqref{bbKRazlLapl})
\begin{multline} \label{K_F^M}
\hat K_{F^M}(\tau| x,x') = \frac{\Delta^{-1}(x,x')}{(4\pi\tau^{1/M})^{d/2}} \, g^{1/2}(x')\\
\times\sum\limits_{m=0}^\infty \tau^{m/M}\,\calE_{M, d/2-m} \left(-\frac{\sigma}{2\tau^{1/M}}\right)\,\hat b^F_m(x,x').
\end{multline}

It is clearly seen that the two expansions are substantially different and cannot be directly compared term by term. Comparison of (\ref{MainExpansionF^Z}) and (\ref{K_F^M}) would only be possible if we additionally expand both expressions in powers of the argument of generalized exponential functions $-\sigma/2\tau^{1/M}$. This in its turn would correspond to double covariant Taylor series in {\em two independent} variables $\tau$ and $\sigma^{a'}(x,x')/\tau^{1/2M}$. The coefficients of these series are, obviously, the coincidence limits of multiple covariant derivatives $\nabla_{a_1}\cdots\nabla_{a_n}\hat K(\tau|x,x')\,|_{\,x'=x}$---exactly the ones that were considered in Sect.\ref{Examples} when comparing our results with conventional Schwinger--DeWitt technique.

This, in particular, explains why the result (\ref{K_F^M}) for integer power of the minimal second order operator (with its absence of $m<0$ terms) does not contradict the nonzero value (\ref{NegPow}) of $\hat b_{-1,1}(x,x')$ for the fourth order operator (\ref{Gen4Order}). Even if one adjusts its coefficients $\hat\varOmega^{abc}$, $\hat D^{ab}$, $H^a$ and $\hat P$ so that the operator becomes the square of the second order minimal one, its $\hat b_{-1,1}(x,x')$ does not identically vanish because of the first term of \eqref{NegPow}---this is a simple demonstration that term by term comparison of (\ref{MainExpansionF^Z}) and (\ref{K_F^M}) does not work.

\section{Discussion and conclusions} \label{Conclusion}

We have suggested a systematic calculational method for the covariant expansion of the two-point heat kernel for generic minimal and non-minimal differential operators of any order. This expansion for off-diagonal heat kernel elements is not an expansion in positive powers of the proper time, like it happens for the minimal second order operators within a conventional Schwinger--DeWitt technique. Rather this is an expansion in background dimensionality of relevant background field objects describing the dimensionful coefficients of the operator and the corresponding spacetime and vector bundle curvatures. Any order of this expansion in powers of these curvatures and their covariant derivatives can be reached in a finite number of calculational steps, which is what one usually needs in local gradient expansion in the energy domain below the effective field theory cutoff. For the coincidence limit of the heat kernel and its arbitrarily high covariant derivatives this expansion becomes the series in positive fractional powers of the proper time. This makes our results fully consistent with the structure of asymptotic expansion of the functional trace of the heat kernel well known from mathematics literature on differential and pseudo-differential operators on curved manifolds \cite{Seeley, Gilkey1975, Gilkey1979}.

Main results of the paper for generic non-minimal operators of order $N$ are given by Eqs.(\ref{bbKRazl})-(\ref{IntL3}), where the set of auxiliary differential operators $\hat T_{n,l}(\nabla)$ is defined by the recurrence relations (\ref{FrakTRecRel}). The efficiency of this result depends on the users skill of calculating the momentum space integral (\ref{IntL3}) for a given principal symbol of the operator (\ref{symbol}). For general minimal operators, whose principal term represents $M$-th power of Laplacian, this integral can be universally calculated in terms of generalized exponential functions  (\ref{calEInt}) and symmetric tensors (\ref{IntL5}). This leads to the expansion (\ref{MainExpansion})-(\ref{N_m}) with the generalized HaMiDeW coefficients $\hat b_{m,n}(x,x')$ defined by Eqs.(\ref{bmn})-(\ref{Rnm}).

There are two basic differences of this expansion from the $M=1$ case of Schwinger-DeWitt technique. Firstly, one overall exponential function of the argument $-\sigma(x,x')/2\tau^{1/M}$  gets replaced by various generalized exponential functions of the same argument, multiplying individual fractional powers of the proper time. Secondly, fractional powers of the proper time series extend to minus infinity. In the coincidence limit $x'=x$, however, the coefficients of the negative powers vanish in full accordance with the known properties of the heat kernel trace \cite{Seeley, Gilkey1975, Gilkey1979}.

Note that the main recursive ingredient of the heat kernel technique is the solution of the chain of relations (\ref{FrakTRecRel}). Even though the anticipation of the ``bottomless'' chain of HaMiDeW coefficients for higher-derivatives operators, expressed in \cite{Fulling}, is correct, the recursive procedure for them turns out to be possible. The structure of these recurrent equations is more complicated as compared to the Schwinger-DeWitt expansion, but they can be successively solved. Their formulation crucially depends on the decomposition (\ref{ExpF1}) of the original operator into the auxiliary operators $\lb\hat F\rb_m$ and $\lb\hat F\rb_{m,n}$ introduced in Sect.\ref{Fourier}.

Complexity of the minimal higher derivative operators manifests itself in the double series nature of the expansion (\ref{MainExpansion}). As shown in Sect. \ref{PowerMin}, for minimal second order operators double series reduces  with the aid of Eqs.(\ref{Prefactor}) and (\ref{LaplaceBcoeficients}) to the original expansion with the DeWitt coefficients $\hat a_m(x,x')$ composed from their generalized double indexed version $\hat b_{m,n}(x,x')$. In fact, Eq.(\ref{Prefactor}) constitutes an alternative representation of the off-diagonal Schwinger-DeWitt coefficients $\hat a_m(x,x')$. This representation in terms of the infinite series (\ref{LaplaceBcoeficients}) of the $\hat b_{m,n}$ coefficients is not a priori equivalent to the covariant Taylor expansion of $\hat a_m(x,x')$. Thus, it might perhaps open the prospects of their explicit (not recursive) calculation and as a byproduct partial resummation of Schwinger-DeWitt series, discussed from a somewhat other viewpoint in \cite{Ivanov2019, Ivanov2019uxu}. This resummation might be facilitated by the fact that certain subsets of the operators $\hat T_{n,l}(\nabla)$, participating in the construction of $\hat b_{m,n}(x,x')$ coefficients, can be obtained beyond recursive procedure, see Eqs.(\ref{special_T}), (\ref{UltraMarginal}) and (\ref{special_b}).

It should be emphasized that our results for two-point heat kernels are essentially of local nature. In particular, for a general curved spacetime the transition to the locally geodetic coordinate system of $\sigma^{a'}(x',x)$ centered at the point $x'$ is restricted by the vicinity of this point where $\Delta(x,x')$ is nonsingular. This domain is bounded by the caustic points $x$ conjugated to $x'$ with $\Delta(x,x')=\infty$, at which the relation (\ref{Prefactor}) becomes singular. For practical purposes of UV renormalization this restriction is not very important, because the coincidence limit of the heat kernel and all its derivatives is basically everything what one needs in effective field theory. In this respect, however, it is quite interesting that the generalized Fourier method does not recover in the heat kernel expansion (\ref{MainExpansion}) a one-half power of $\Delta(x,x')$ typical for the Schwinger-DeWitt expansion (\ref{HeatKernel0}). Instead it gives the inverse of $\Delta(x,x')$. Thus, naively this would mean that approaching the caustic gives a decreasing amplitude of the physical signal rather than its enhancement which is characteristic of semiclassical approximation.

The explanation of this paradox apparently consists in the observation that for higher derivative operators with $M>1$ our Fourier method is critically different from the semiclassical expansion. If one calculates the momentum space integrals (\ref{calEInt}) by the saddle point method for $\tau\to 0$ then one would get the expansion in powers of $\tau^{1/(2M-1)}$ rather than $\tau^{1/M}$ and, moreover, the terms of this expansion will be singular at $\sigma(x,x')\to 0$ \cite{Wach2}. For $M>1$ the semiclassical asymptotic expansion at $\tau\to 0$ is not homogeneous for $x'\to x$ but conveys a right behavior for large $\sigma(x,x')$ even when approaching the caustic. This, in particular, does not allow us to use the semiclassical approximation
method of Maslov and Fedoriuk \cite{MaslovFedoriuk} in this limit. Conversely, our background dimensionality expansion is regular at $x\to x'$, but fails for large distances. Interestingly, for minimal operators with $M=1$ these two expansions look equivalent in the vicinity of the heat kernel diagonal. Here $\Delta(x,x')$ is close to one, and in the relation (\ref{Prefactor}) only affects nontrivial match between multiple covariant derivatives of $\hat a_m(x,x')$, $\Delta(x,x')$ and $\hat b_m(x,x')$ what was explicitly checked in Sect.\ref{Examples}.

Local nature of our method can be scrutinized also from another viewpoint. Obviously, the expansion up to infinitely positive and negative powers of $\tau$ cannot be regarded as asymptotic for $\tau\to 0$. But the infinite tail of negative powers of $\tau$ should be regarded as an expansion in {\em positive} powers of the ratio $\sigma^{a'}(x,x')/\tau^{1/2M}$, so that the total result is the sum of two expansions which are both asymptotic at $\tau\to 0$ in the domain $\sigma^{a'}(x,x')\ll\tau^{1/2M}$. The coefficients of both expansions are local curvature invariants of growing dimensionality, which explains the efficiency of this method in effective field theory and in renormalization theory.

One might notice that on top of very high generality of the suggested expansions and their detailed analysis no new results were obtained in concrete applications of this technique. This can be explained by the fact that any step beyond second order minimal operators is practically impossible without using powerful symbolic manipulating computer programs. Even the simplest examples of Sect.\ref{Examples}, focused on the comparison of the new method with well known calculations of the lowest orders of the Schwinger--DeWitt expansion, would be impossible without using a computer code. While the solution of DeWitt's recurrent equations for $\hat a_m(x,x)$ with small $m=0,1,2,3$ can be easily performed manually, the same calculation by means of generalized DeWitt coefficients $\hat b_{m,n}(x,x)$ turns out to be much more complicated, and it was accomplished above only by using the xAct package of Mathematica. Extension beyond the case of $M=1$ minimal operators is even more complicated and, in fact, is impossible without the algorithms obtained above. This is undoubtedly true modulo exceptional cases and special tools, like the method of universal functional traces \cite{JackOsborn,Barvinsky85}, the method of squaring the low-derivative operator, dimensional reduction \cite{Barvinsky2021ubv}, etc. So the ultimate goal of this work is to develop on the basis of above algorithms an efficient code for calculating the heat kernel expansion and using it in high energy applications.

One of the motivations for this project is the problem of UV renormalization in Ho\v{r}ava gravity theory, which is computationally extremely complicated because of Lorentz invariance violation. Extension of the heat kernel method to Lorentz violating models is also possible \cite{Solodukhin2011, Barvinsky172, Saueressig_etal} and includes a recent application of the Fourier method \cite{Grosvenor2021zvq} along the lines of the Gusynin method for the operator resolvent \cite{Gusynin1989, Gusynin1990}. In four
dimensions this problem involves non-minimal operators of the sixth order \cite{Barvinsky2021ubv} and requires the whole set of special methods involving the above mentioned universal functional traces \cite{Barvinsky85}, 3-dimensional reduction on a static background, square root extraction for the sixth-order operator having hundreds of terms, etc. At present these are the only available tools of the background field method needed to avoid calculation of humongous number of Feynman graphs, that was, for example, undertaken in the proof of asymptotic freedom in (2+1)-dimensional Ho\v{r}ava gravity \cite{Barvinsky172}. However, these auxiliary tools are not always possible and often lengthen the needed computations like it happens with the method of \cite{Barvinsky2021ubv}, which relies on Sylvester procedure for extracting the square root of the sixth-order operator in covariant derivatives. The presented technique can and should circumvent these difficulties, and the comparison of its efficiency with the approach of \cite{Barvinsky2021ubv} is one of our main future goals.

We accomplish the paper with a couple of brief remarks on possible future refinement of the suggested formalism. The failure to perform a complete resummation of negative powers of the proper time suggests that such a resummation is perhaps, in principle, impossible. This might be explained by the fact that negative powers of $\tau$ entail positive powers of the vector $\sigma^a(x,x')$ rather than the scalar function $\sigma(x,x')$. This conclusion seems to be corroborated by expressions like \eqref{NegPow}. Therefore the coefficients of this expansion are tensors of growing rank, and this invalidates universality of the needed resummation, leaving however the option of summing up the series of particular tensor structures.

Another remark is that our Fourier method consists of multiple stages---construction of auxiliary operators $\lb F\rb_{m,n}$ in the decompositions (\ref{decomp1}) and (\ref{Commut}), recursive solution of equations for the operators $\hat T_{n,l}(\nabla)$, necessity of finding coincidence limits of multiple derivatives of $\sigma$, presence of primed and unprimed indices, etc. All this essentially slows down symbolic computations. This suggests the necessity of an alternative approach combining the advantages of the commutator technique of the universal functional traces of \cite{Barvinsky85,Barvinsky2021ubv} and the above Fourier method. Such approach indeed can be worked out, and it is anticipated that in the new method all generalized heat kernel coefficients $\hat b_{m,n}(x,x')$ could be expressed in terms of the conventional DeWitt coefficients $\hat a_m^{(-\Box)}(x,x')$ of the simplest minimal operator $-\Box$. This can be achieved via acting upon them by the set of differential operators which are formed of the products of multiple commutators of $(-\Box)^M$ with the lower derivative part $\hat P(\nabla)$ of the full minimal operator (\ref{Min_O}),
\begin{equation}
[(-\Box)^M,[(-\Box)^M,...[(-\Box)^M,\hat P(\nabla)],...]],
\end{equation}
This is the idea of the method of universal functional traces \cite{Barvinsky85, Barvinsky2021ubv}, exploiting a simple fact that every commutation with $\Box$ increases the dimensionality of any object at least by one. This algorithm, which is currently under study \cite{Work_in_progress}, is expected to be especially suitable for the needs of effective field theory, because it clearly shows how a needed precision of curvature expansion truncates the orders of multiple commutators.

\section*{Acknowledgments}
The authors are deeply grateful to Alexander Kurov for his assistance in computer symbolic calculations. We are also indebted to Darren Grasso for the correction of matrix algebra in the case of generic non-minimal operators. The work was partially supported by the Russian Foundation for Basic Research grant 20-02-00297 and by the Foundation for Theoretical Physics Development ``Basis''.

\appendix
\section{The world function and the parallel transport tensor } \label{AppA}

First of all agree on basic definitions and notation. Let us be given some $d$-dimensional manifold $\calM$, which is the base of the vector bundle $\pi\colon \calE\to\calM$. We will use lowercase Latin letters for spacetime indices, and we will usually omit indices in the bundle, denoting endomorphisms in fibers by hats. Further, let the connection $\nabla_a$ be given on the sections of this bundle. We everywhere assume that this connection is symmetric, so that there is no torsion $T^c{}_{ab} = 0$ and covariant derivatives commute on scalars. Then the Riemann tensor and curvature in the bundle are defined as
\begin{gather}
[\nabla_a, \nabla_b] v^c = R^c{}_{dab} v^d, \label{DefR} \\
[\nabla_a, \nabla_b] \varphi = \hat\calR_{ab} \varphi. \label{DefF}
\end{gather}
For the action of $[\nabla_a, \nabla_b]$ on a matrix-valued function, we obtain
\begin{equation}
[\nabla_a, \nabla_b] \hat P = [\hat\calR_{ab}, \hat P].
\end{equation}
The convolution of the Riemann tensor with respect to a pair of indices determine the Ricci tensor
\begin{equation}
R_{ab} = R^c{}_{acb}.
\end{equation}

On a manifold with connection, we can introduce geodesics, as lines, the tangent vector to which goes into itself by a parallel transport along them. This leads to the following definition of tangent vectors $\sigma^a(x,x')$
\begin{equation} \label{SigmaDef}
\sigma^b\nabla_b\sigma^a=\sigma^a, \qquad \big[\,\sigma^a \big] = 0.
\end{equation}
Similarly, we can transport along geodesics a tensor of arbitrary nature with an index in the bundle. The condition on the parallel transport tensor $\hat\calI(x,x')$ will have the form
\begin{equation} \label{calIDef}
\sigma^a\nabla_a\hat\calI = 0, \qquad \big[\,\hat\calI\, \big] = \hat 1.
\end{equation}

For the derivatives of $\sigma^a(x,x')$ and $\hat\calI(x,x')$, we introduce the following abbreviations
\begin{align}
\sigma^a{}_{b_1\ldots b_n} &= \nabla_{b_n}\cdots\nabla_{b_1}\sigma^a, \\
\hat\calI_{b_1\ldots b_n} &= \nabla_{b_n}\ldots\nabla_{b_1}\hat\calI.
\end{align}
Then the definition \eqref{SigmaDef} takes the form
\begin{equation}
\sigma^a{}_b \sigma^b = \sigma^a. \label{Projection}
\end{equation}
Differentiating \eqref{Projection}, we get $\sigma_a{}^b \sigma_b{}^c + \sigma_a{}^{bc} \sigma_b = \sigma_a{}^c$, which in the coincidence limit leads to
\begin{equation}
[\sigma^a{}_b] = \delta_b^a.
\end{equation}

Sequentially acting on the relations \eqref{Projection} and \eqref{calIDef} by the operator $\sigma^c\nabla_c$ and simplifying the expressions using \eqref{Projection}, we obtain the important relations
\begin{gather}
\sigma^a{}_{b_1\ldots b_n} \sigma^{b_1} \ldots \sigma^{b_n} = 0, \quad n\ge2, \label{SigmaSymm} \\
\sigma^{a_1}\ldots\sigma^{a_k} \hat\calI_{a_1\ldots a_k} = 0.
\end{gather}

We emphasize once again that the introduction of geodesics and related objects $\sigma^a(x,x')$ and $\hat\calI(x,x')$ and their properties does not require the existence of the metric $g_{ab}(x)$ (although, of course, in this case there is no natural parameter along the geodesic).

If $\calM$ is a Riemannian manifold with metric $g_{ab}(x)$, then the conditions of covariant constancy of the metric $\nabla_a g_{bc}=0$ determine the Christoffel connection
\begin{gather}
\nabla_a v^c = \partial_a v^c + \Gamma^c{}_{ab} v^b, \\
\Gamma^c{}_{ab} = \frac{1}{2} g^{cd} \left(\partial_a g_{bd} + \partial_b g_{ad} - \partial_d g_{ab}\right).
\end{gather}
The convolution of the Ricci tensor determines the scalar curvature
\begin{equation}
R = R^a{}_a.
\end{equation}

The geodesics divergence rate $\sigma_a{}^a$ is closely related with the Pauli--Van Vleck--Morette determinant
\begin{gather}
\Delta(x,x') = \frac{\det\left(-\nabla_a\nabla_{b'}\sigma\right)}{g^{1/2}(x)\,g^{1/2}(x')}, \label{Van_Vleck} \\
\sigma^a\nabla_a \Delta = \Delta (d - \sigma_a{}^a).
\end{gather}
If the point $x$ lies on the caustic of the point $x'$ (that is, if the geodesics outgoing from $x'$ intersect at $x$), the determinant $D(x,x')$ becomes infinite. However, we are everywhere interested only in the local (ultraviolet) properties of the quantities under consideration, i.e. the case when $x$ and $x'$ are close enough to each other.

Finally, we can calculate the coincidence limits $[\sigma_{a_1\cdots a_n}]$ and $[\hat\calI_{a_1\cdots a_n}]$. To do this, one needs to differentiate \eqref{Projection} and \eqref{calIDef} the necessary number of times, go to the coincidence limits in the expressions obtained, and order the derivatives in them using \eqref{DefR}--\eqref{DefF}. As a result, we get
\begin{align}
[\sigma_{abc}] &= 0, \\
[\sigma_{a_1a_2a_3a_4}] &= \frac{2}{3}R_{a_1(a_3a_4)a_2}, \\
[\sigma_{a_1a_2a_3a_4a_5}] &= \frac{3}{2} \nabla_{(a_3} R_{|a_1|a_4a_5)a_2}, \\
[\hat\calI_a] &= 0, \\
[\hat\calI_{ab}] &= \frac{1}{2} \hat\calR_{ab}, \\
[\hat\calI_{abc}] &= \frac{2}{3} \nabla_{(a}\hat\calR_{b)c}
\end{align}
and so on.

\section{The generalized exponential functions (GEF)} \label{GEF}

In the articles \cite{Wach, Wach2} we showed that the heat kernel of a power of the Laplacian $(-\Box)^\nu$ in a flat $d$-dimensional space with the Euclidean metric has the form
\begin{multline} \label{FunkEvol}
\bbK_{\nu,d}(\tau; x) = \exp\left(-\tau(-\Box)^\nu\right) \delta(x) \\
= \frac1{\big(4\pi\tau^{1/\nu}\big)^{d/2}}\, \calE_{\nu, d/2}\left(-\frac{|x|^2}{4\tau^{1/\nu}}\right),
\end{multline}
where $\calE_{\nu,\alpha}(z)$ is a two-parameter family of some new special functions defined by the Taylor series
\begin{equation}
\calE_{\nu,\alpha}(z) = \frac{1}{\nu}\sum\limits_{m=0}^\infty \frac{\Gamma\left(\frac{\alpha+m}{\nu}\right) }{\Gamma(\alpha+m)} \frac{z^m}{m!}.
\end{equation}
In particular
\begin{equation} \label{calE0}
\calE_{\nu,\alpha}(0) = \frac{\Gamma(\alpha/\nu)}{\nu\Gamma(\alpha)}.
\end{equation}

The expression \eqref{FunkEvol} is a direct generalization of the standard heat kernel for the Laplacian $-\Box$ for $\nu=1$
\begin{gather} \label{HeatKernel}
\calE_{1,\alpha}(z) = \exp(z), \\
\bbK_{1,d}(\tau; x) = \frac{1}{(4\pi\tau)^{d/2}} \exp\left(-\frac{|x|^2}{4\tau}\right),
\end{gather}
in which the functions $\calE_{\nu,d/2}(z)$ are in place of the usual exponential functions. So we call these new functions \emph{the generalized exponential functions (GEF)}. We investigated their properties in detail in \cite{Wach, Wach2}. The most important of them are the direct and inverse Mellin transform and the differentiation rule
\begin{align}
&\varepsilon_{\alpha,a}(s) \equiv \int\limits_0^\infty z^{s-1} \calE_{\alpha,a}(-z) dz = \frac{\Gamma(s)\Gamma\left(\frac{a-s}{\alpha}\right)}{\alpha\Gamma(a-s)}, \label{MellinCalE} \\
&\calE_{\alpha,a}(-z) = \frac{1}{2\pi i} \int\limits_C \varepsilon_{\alpha,a}(s) z^{-s} ds, \label{InvMellin} \\
&\frac{d^b}{dz^b} \calE_{\alpha,a}(z) = \calE_{\nu,a+b}(z). \label{DiffRule}
\end{align}

Let us also introduce new functions which we will call \emph{the GEF of the second order}
\begin{gather}
\varepsilon_{\nu,b}^{\alpha,a}(s) = \frac{\Gamma(s) \Gamma\left(\frac{a-s}{\alpha\nu}\right) \Gamma\left(\frac{b-s}{\nu}\right)}{\alpha\nu \Gamma\left(\frac{a-s}{\nu}\right)\Gamma(b-s)}, \label{epsilon2order}\\
\calE_{\nu,b}^{\alpha,a}(-z) = \frac{1}{2\pi i} \int\limits_C \varepsilon_{\nu,b}^{\alpha,a}(s)\, z^{-s}\, ds.     \label{GEF2order}
\end{gather}
Their properties are:
\begin{gather}
\calE_{1,b}^{\alpha,a}(z) = \calE_{\alpha,a}(z), \qquad \calE_{\nu,b}^{1,a}(z) = \calE_{\nu,b}(z), \\
\calE_{\nu,b}^{\alpha,a}(z) = \sum\limits_{n=0}^\infty \frac{\Gamma\left(\frac{a+n}{\alpha\nu}\right) \Gamma\left(\frac{b+n}{\nu}\right)}{\alpha\nu \Gamma\left(\frac{a+n}{\nu}\right) \Gamma(b+n)} \frac{z^n}{n!}.
\end{gather}
In particular
\begin{equation} \label{H0}
\calE_{\nu,b}^{\alpha,a}(0) = \frac{\Gamma(a/\alpha\nu) \Gamma(b/\nu)}{\alpha\nu \Gamma(a/\nu) \Gamma(b)}.
\end{equation}

On account of these properties of GEF and second order GEF, the derivation of Eq.(\ref{bbKn}) consists in the change of integration variable $t\to\mu = d/2 - m - Mst$,
\begin{multline} \label{bbKnA}
\frac{1}{2\pi i} \int\limits_{w-i\infty}^{w+i\infty} dt\,\tau^{-t}\, \Gamma(t)\, \bbG^{m,n}(st,\sigma)\, \\
= \frac{\tau^{m/sM}}{(4\pi\tau^{1/sM})^{d/2}}\int\limits_C \frac{d\mu}{2\pi i}\,z^{-\mu}\, \varepsilon_{M, d/2+Mn-m}^{s, d/2-m}(\mu),
\end{multline}
and the use of Eqs.(\ref{epsilon2order})-(\ref{GEF2order}) with the argument $z = \sigma/2\tau^{1/\alpha M}$.

\section{Complex powers} \label{GeneralTheory}

The complex power $F^{-s}$ of the operator $F$ for $s\ne 0,-1,-2, \ldots$ can be formally defined as the following integral of the operator $e^{-\tau F}$
\begin{equation} \label{FNu}
F^{-s} = \frac{1}{\Gamma(s)}\int\limits_0^\infty d\tau\, \tau^{s-1} e^{-\tau\hat F}.
\end{equation}
For $s\in\mathbb{N}$ the relation (\ref{FNu}) can be verified by alternately acting on it by the operator $F$ and integrating by parts $s$ times. Note that from the definition \eqref{FNu}, using the properties of gamma functions, it is not difficult to obtain properties that are naturally expected from complex powers, for example, $F^a\,F^b = F^{a+b}$.

The inverse transform is given by the formula
\begin{equation}
e^{-\tau F} = \frac{1}{2\pi i} \int\limits_{w-i\infty}^{w+i\infty} \frac{\tau^{-s} \Gamma(s)}{ F^s} ds,
\end{equation}
The last expression is easy to understand as follows: the function $\Gamma(s)$ has simple poles at the points $s_n = -n$ with residues $(-1)^n/n!$. Then the integral over $s$ reduces to the sum of the residues at these poles, which exactly gives the standard Taylor series expansion for the operator exponent $e^{-\tau F}$.

Note that mathematics of spectral geometry and related topics is usually limited to compact manifolds, which leads to significant simplifications associated with the discrete spectrum of $F$. Indeed, in this case, for a strictly positive elliptic differential operator $F=\hat F(\nabla)$, there exists a discrete orthonormal basis of eigenfunctions
\begin{equation}
|n\rangle = \phi_n(x), \qquad  F |n\rangle = \lambda_n |n\rangle, \qquad \lambda_n>0.
\end{equation}

Then the operator itself and functions of it can be expressed in a simple way in terms of eigenfunctions and eigenvalues
\begin{gather}
F = \sum_n \lambda_n |n\rangle \langle n|, \\
e^{-\tau F} = \sum_n e^{-\tau\lambda_n} |n\rangle \langle n|, \\
\frac{1}{ F + \lambda} = \sum_n \frac{|n\rangle \langle n|}{\lambda_n + \lambda}, \\
F^{-s} = \sum_n \lambda_n^{-s} |n\rangle \langle n|.
\end{gather}
Using the functional trace allows to determinate important global quantities such as the heat kernel trace and the standard operator zeta function
\begin{gather}
\Tr e^{-\tau F} = \sum_n e^{-\tau\lambda_n}, \\
\zeta_F(s) = \Tr F^{-s} = \sum_n \lambda_n^{-s}.
\end{gather}
On compact manifolds these global quantities can be obtained by the integration of local ones
\begin{gather}
\Tr e^{-\tau\hat F} = \int d^dx\, g^{1/2}(x)\, \tr\hat K_F(\tau|x,x).
\end{gather}
For non-compact manifolds and manifolds with boundaries, which are important for physical applications, the definition of global quantities requires additional care because of boundary conditions. The difficulties associated with this can be circumvented by working directly with local (i.e., point-dependent) quantities which asymptotically in small $\tau$ and $\sigma^a/\tau^{1/2M}$ are not sensitive to the presence of boundaries. This is the case we consider in this paper.

\section{$\varOmega$-terms \label{Omega}}
Additional contributions $\hat E_2^\varOmega$ and $\hat E_4^\varOmega$ of nonzero $\hat\varOmega^{abc}$ to the Gilkey-Seeley coefficients $\hat E_2$ and $\hat E_4$ of the minimal fourth-order operator (\ref{Gen4Order}) can be relatively concisely written down with the aid of the totally symmetric tensor
\begin{equation}
g_{a_1\cdots a_{2n}} = [S_{n,2n}] = \frac{(2n)!}{2^n n!} g_{(a_1a_2}\cdots g_{a_{2n-1}a_{2n})},
\end{equation}
where symmetrization is taken over all $2n$ indices (with the coefficient $1/(2n)!$).

The expression for $\hat E_2^\varOmega$ is rather simple
\begin{multline}
\hat E_2^\varOmega = - \frac{1}{(4\pi)^{d/2}} \frac{3\,\Gamma\left(\frac{d/2-1}{2}\right)}{8\,d\,\Gamma(\frac{d}2-1)}\\
\times\Big\{ \frac{1}{4(d+4)}\Big(2\hat\varOmega_{abc}\hat\varOmega^{abc} + 3\hat\varOmega_a\hat\varOmega^a\Big) + \nabla_a\hat\varOmega^a \Big\},
\end{multline}
whereas for $\hat E_4^\varOmega$ it is a sum of the contribution of $\hat\varOmega^{abc}$ alone and the cross terms of $\hat\varOmega^{abc}$ with Ricci curvature, fibre bundle curvature and the coefficients of the operator $\hat D^{ab}$ and $\hat H^a$,
\begin{equation}
\hat E_4^\varOmega = \frac{1}{(4\pi)^{d/2}} \frac{\Gamma\left(\frac{d}{4}\right)}{2\Gamma\left(\frac{d}{2}\right)} \Big\{ \hat B_\varOmega + \hat B_{\varOmega R} + \hat B_{\varOmega\calR} + \hat B_{\varOmega DH}\Big\}.
\end{equation}

Separately these contributions look as follows. The most complicated is the first term $\hat B_\varOmega$ which is represented as a quartic polynomial in $\hat\varOmega^{abc}$ and its derivatives,
\begin{widetext}
\begin{equation}
\begin{split}
&\hat B_\varOmega= -\frac{1}{8}(\nabla_a\Box + \Box\nabla_a)\hat\varOmega^a + \frac{d}{8(d+2)} (\nabla_b\nabla_a\nabla^b\hat\varOmega^a + 2\nabla_a\nabla_b\nabla_c\hat\varOmega^{abc})  \\
&\qquad+ \frac{1}{32(d+2)(d+6)}\bigg\{3(d+4)
\Big(2\hat\varOmega^{abc}\nabla_a\nabla_b\hat\varOmega_c
+ 2(\nabla_a\nabla_b\hat\varOmega_c)\hat\varOmega^{abc} + 2(\nabla_c\nabla^d\hat\varOmega_{abd})\hat\varOmega^{abc}
+ 2(\nabla^d\nabla_c\hat\varOmega_{abd})\hat\varOmega^{abc}
\\
&\qquad+ 2(\nabla_a\nabla_b\hat\varOmega^{abc})\hat\varOmega_c+ (\nabla_b\nabla_a\hat\varOmega^b)\hat\varOmega^a - (\nabla_a\nabla_b\hat\varOmega^b)\hat\varOmega^a\Big)
+ (d+8) \Big(-3\hat\varOmega^a\nabla_a\nabla_b\hat\varOmega^b  - 6(\nabla_a\hat\varOmega^{abc})\nabla_b\hat\varOmega_c
\\
&\qquad+ 6\hat\varOmega^{abc}\nabla_c\nabla^d\hat\varOmega_{abd}
+ 3(\nabla_a\hat\varOmega_b)\nabla^b\hat\varOmega^a + 3(\nabla_a\hat\varOmega_b)\nabla^a\hat\varOmega^b - 3\hat\varOmega^a\nabla_b\nabla_a\hat\varOmega^b - 6\hat\varOmega^{abc}\nabla^d\nabla_c\hat\varOmega_{abd} - 2\hat\varOmega_{abc}\Box\hat\varOmega^{abc}
\\
&\qquad - 3\hat\varOmega_a\Box\hat\varOmega^a - 2(\Box\hat\varOmega_{abc})\hat\varOmega^{abc} - 3(\Box\hat\varOmega_{abc})\hat\varOmega^{abc} - 6(\nabla_d\hat\varOmega_{abc})\nabla^c\hat\varOmega^{abd}
+ 2(\nabla_d\hat\varOmega_{abc})\nabla^d\hat\varOmega^{abc}\Big)
\\
& \qquad+ 3(3d+16)\Big(2(\nabla_a\hat\varOmega^{acd})\nabla^b\hat\varOmega_{bcd} + 2(\nabla_a\hat\varOmega_b)\nabla_c\hat\varOmega^{abc} + (\nabla_a\hat\varOmega^a)^2 + 2\hat\varOmega_a\nabla_b\nabla_c\hat\varOmega^{abc}\Big)  \bigg\}
\\
&\qquad- \frac{g_{abcdefgh}}{32(d+2)(d+6)} \Big(\hat\varOmega^{iab}\hat\varOmega^{cde}\nabla_i\hat\varOmega^{fgh} + \hat\varOmega^{abc}\hat\varOmega^{ide}\nabla_i\hat\varOmega^{fgh} + \hat\varOmega^{iab}(\nabla_i\hat\varOmega^{cde})\hat\varOmega^{fgh}\Big)
\\
&\qquad+ \frac{g_{abcdefghij}}{384(d+2)(d+6)} \Big((\nabla^a\hat\varOmega^{bcd})\hat\varOmega^{efg}\hat\varOmega^{hij} + 2\hat\varOmega^{abc}(\nabla^d\hat\varOmega^{efg})\hat\varOmega^{hij} + \hat\varOmega^{abc}\hat\varOmega^{def}\nabla^g\hat\varOmega^{hij}\Big)
\\
&\qquad+\frac{g_{abcdefghijkl}}{1536(d+2)(d+6)(d+10)}
\hat\varOmega^{abc}\hat\varOmega^{def}\hat\varOmega^{ghi}\hat\varOmega^{jkl}.
\end{split}
\end{equation}

The cross terms correspondingly read
\begin{align}
&\hat B_{\varOmega R} = -\frac{1}{8} R\nabla_a\hat\varOmega^a + \frac{1}{4}R_{ab}\nabla_c\hat\varOmega^{abc}
+ \frac{1}{32(d+2)} \Big(6R_{ab}\hat\varOmega^a\hat\varOmega^b + 6R_{ab}\hat\varOmega_c\hat\varOmega^{abc} - 3R\hat\varOmega_a\hat\varOmega^a - 2R\hat\varOmega_{abc}\hat\varOmega^{abc}
\nonumber\\
&\qquad+ 4\hat\varOmega^{abc}\nabla_aR_{bc}+ 2\hat\varOmega^a\nabla_aR + 8R_{ab}\nabla_c\hat\varOmega^{abc}\Big),\\
&\hat B_{\varOmega\calR} = \frac{1}{16(d+2)} \Big(4(d+4)(\nabla_b\hat\varOmega_a)\hat\calR^{ab} - 8\hat\varOmega_b\nabla_a\hat\calR^{ab} - 6\hat\varOmega_{acd}\hat\varOmega_b{}^{cd}\hat\calR^{ab} - 3\hat\varOmega_a\hat\varOmega_b\hat\calR^{ab}\Big),
\end{align}
\begin{align}
&\hat B_{\varOmega DH} = \frac{3}{8(d+2)} \Big(\hat H_a\hat\varOmega^a + \hat\varOmega^a\hat H_a\Big) + \frac{1}{8(d+2)}\Big(-4\hat\varOmega_a\nabla_b\hat D^{ab} - 2(\nabla_a\hat D^{ab})\hat\varOmega_b + 2\hat D^{ab}\nabla_a\hat\varOmega_b - 2(\nabla_a\hat\varOmega_b)\hat D^{ab}
\nonumber \\
&\qquad+ \hat\varOmega^a\nabla_a\hat D - (\nabla_a\hat D)\hat\varOmega^a - 4\hat D_{ab}\nabla_c\hat\varOmega^{abc} - 2 \hat D\nabla_a\hat\varOmega^a - 2 (\nabla_a\hat\varOmega^{abc})\hat D_{bc} - (\nabla_a\hat\varOmega^a)\hat D + 2\hat\varOmega^{abc}\nabla_a\hat D_{bc} - 2(\nabla_a\hat D_{bc})\hat\varOmega^{abc}\Big)
\nonumber\\
&\qquad- \frac{g_{abcdefgh}}{96(d+2)(d+6)} \Big(\hat D^{ab}\hat\varOmega^{cde}\hat\varOmega^{fgh} + \hat\varOmega^{abc}\hat D^{de}\hat\varOmega^{fgh} + \hat\varOmega^{abc}\hat\varOmega^{def}\hat D^{gh}\Big).
\end{align}
\end{widetext}
Here $\hat\varOmega^a$ and $\hat D$ are given by the contractions with the metric
\begin{equation}
\hat\varOmega^a=g_{bc}\hat\varOmega^{abc},\quad \hat D=g_{ab}\hat D^{ab}.
\end{equation}

\bibliography{HeatKernel}

\begin{thebibliography}{44}%
\makeatletter
\providecommand \@ifxundefined [1]{%
 \@ifx{#1\undefined}
}%
\providecommand \@ifnum [1]{%
 \ifnum #1\expandafter \@firstoftwo
 \else \expandafter \@secondoftwo
 \fi
}%
\providecommand \@ifx [1]{%
 \ifx #1\expandafter \@firstoftwo
 \else \expandafter \@secondoftwo
 \fi
}%
\providecommand \natexlab [1]{#1}%
\providecommand \enquote  [1]{``#1''}%
\providecommand \bibnamefont  [1]{#1}%
\providecommand \bibfnamefont [1]{#1}%
\providecommand \citenamefont [1]{#1}%
\providecommand \href@noop [0]{\@secondoftwo}%
\providecommand \href [0]{\begingroup \@sanitize@url \@href}%
\providecommand \@href[1]{\@@startlink{#1}\@@href}%
\providecommand \@@href[1]{\endgroup#1\@@endlink}%
\providecommand \@sanitize@url [0]{\catcode `\\12\catcode `\$12\catcode
  `\&12\catcode `\#12\catcode `\^12\catcode `\_12\catcode `\%12\relax}%
\providecommand \@@startlink[1]{}%
\providecommand \@@endlink[0]{}%
\providecommand \url  [0]{\begingroup\@sanitize@url \@url }%
\providecommand \@url [1]{\endgroup\@href {#1}{\urlprefix }}%
\providecommand \urlprefix  [0]{URL }%
\providecommand \Eprint [0]{\href }%
\providecommand \doibase [0]{https://doi.org/}%
\providecommand \selectlanguage [0]{\@gobble}%
\providecommand \bibinfo  [0]{\@secondoftwo}%
\providecommand \bibfield  [0]{\@secondoftwo}%
\providecommand \translation [1]{[#1]}%
\providecommand \BibitemOpen [0]{}%
\providecommand \bibitemStop [0]{}%
\providecommand \bibitemNoStop [0]{.\EOS\space}%
\providecommand \EOS [0]{\spacefactor3000\relax}%
\providecommand \BibitemShut  [1]{\csname bibitem#1\endcsname}%
\let\auto@bib@innerbib\@empty
\bibitem [{\citenamefont {Schwinger}(1951)}]{Schwinger}%
  \BibitemOpen
  \bibfield  {author} {\bibinfo {author} {\bibfnamefont {J.}~\bibnamefont
  {Schwinger}},\ }\bibfield  {title} {\bibinfo {title} {{On gauge invariance
  and vacuum polarization}},\ }\href {https://doi.org/10.1103/PhysRev.82.664}
  {\bibfield  {journal} {\bibinfo  {journal} {Phys. Rev.}\ }\textbf {\bibinfo
  {volume} {82}},\ \bibinfo {pages} {664} (\bibinfo {year} {1951})}\BibitemShut
  {NoStop}%
\bibitem [{\citenamefont {DeWitt}(1965)}]{DeWitt}%
  \BibitemOpen
  \bibfield  {author} {\bibinfo {author} {\bibfnamefont {B.~S.}\ \bibnamefont
  {DeWitt}},\ }\href@noop {} {\emph {\bibinfo {title} {{Dynamical Theory of
  Groups and Fields}}}}\ (\bibinfo  {publisher} {Gordon and Breach},\ \bibinfo
  {address} {New York},\ \bibinfo {year} {1965})\BibitemShut {NoStop}%
\bibitem [{\citenamefont {Barvinsky}\ and\ \citenamefont
  {Vilkovisky}(1985)}]{Barvinsky85}%
  \BibitemOpen
  \bibfield  {author} {\bibinfo {author} {\bibfnamefont {A.~O.}\ \bibnamefont
  {Barvinsky}}\ and\ \bibinfo {author} {\bibfnamefont {G.~A.}\ \bibnamefont
  {Vilkovisky}},\ }\bibfield  {title} {\bibinfo {title} {{The generalized
  Schwinger--DeWitt technique in gauge theories and quantum gravity}},\ }\href
  {https://doi.org/10.1016/0370-1573(85)90148-6} {\bibfield  {journal}
  {\bibinfo  {journal} {Phys. Rep.}\ }\textbf {\bibinfo {volume} {119}},\
  \bibinfo {pages} {1} (\bibinfo {year} {1985})}\BibitemShut {NoStop}%
\bibitem [{\citenamefont {Barvinsky}()}]{Scholarpedia}%
  \BibitemOpen
  \bibfield  {author} {\bibinfo {author} {\bibfnamefont {A.~O.}\ \bibnamefont
  {Barvinsky}},\ }\bibfield  {title} {\bibinfo {title} {{Heat kernel expansion
  in the background field formalism}},\ }\bibinfo {note} {scholarpedia 10
  (2015) 6, 31644}\BibitemShut {NoStop}%
\bibitem [{\citenamefont {Avramidi}(2000)}]{Avram00}%
  \BibitemOpen
  \bibfield  {author} {\bibinfo {author} {\bibfnamefont {I.~G.}\ \bibnamefont
  {Avramidi}},\ }\href@noop {} {\emph {\bibinfo {title} {{Heat Kernel and
  Quantum Gravity}}}},\ \bibinfo {series} {Lecture Notes in Physics
  Monographs}\ No.~\bibinfo {number} {64}\ (\bibinfo  {publisher}
  {Springer-Verlag},\ \bibinfo {address} {Berlin, Heidelberg},\ \bibinfo {year}
  {2000})\BibitemShut {NoStop}%
\bibitem [{\citenamefont {Seeley}(1967)}]{Seeley}%
  \BibitemOpen
  \bibfield  {author} {\bibinfo {author} {\bibfnamefont {R.~T.}\ \bibnamefont
  {Seeley}},\ }\bibfield  {title} {\bibinfo {title} {{Complex powers of an
  elliptic operator}},\ }in\ \href {https://doi.org/10.1090/pspum/010} {\emph
  {\bibinfo {booktitle} {Singular Integrals}}},\ \bibinfo {series} {Proc.
  Sympos. Pure Math.}, Vol.~\bibinfo {volume} {10}\ (\bibinfo  {publisher}
  {Amer. Math. Soc.},\ \bibinfo {address} {Chicago, Ill},\ \bibinfo {year}
  {1967})\ pp.\ \bibinfo {pages} {288--307}\BibitemShut {NoStop}%
\bibitem [{\citenamefont {Gilkey}(1975)}]{Gilkey1975}%
  \BibitemOpen
  \bibfield  {author} {\bibinfo {author} {\bibfnamefont {P.~B.}\ \bibnamefont
  {Gilkey}},\ }\bibfield  {title} {\bibinfo {title} {{The spectral geometry of
  a Riemannian manifold}},\ }\href {https://doi.org/10.4310/jdg/1214433164}
  {\bibfield  {journal} {\bibinfo  {journal} {J. Differ. Geom.}\ }\textbf
  {\bibinfo {volume} {10}},\ \bibinfo {pages} {601} (\bibinfo {year}
  {1975})}\BibitemShut {NoStop}%
\bibitem [{\citenamefont {Gilkey}(1979)}]{Gilkey1979}%
  \BibitemOpen
  \bibfield  {author} {\bibinfo {author} {\bibfnamefont {P.~B.}\ \bibnamefont
  {Gilkey}},\ }\bibfield  {title} {\bibinfo {title} {{Recursion relations and
  the asymptotic behavior of the eigenvalues of the Laplacian}},\ }\href@noop
  {} {\bibfield  {journal} {\bibinfo  {journal} {Compositio Math.}\ }\textbf
  {\bibinfo {volume} {38}},\ \bibinfo {pages} {201} (\bibinfo {year}
  {1979})}\BibitemShut {NoStop}%
\bibitem [{\citenamefont {Vassilevich}(2003)}]{Vassil03}%
  \BibitemOpen
  \bibfield  {author} {\bibinfo {author} {\bibfnamefont {D.~V.}\ \bibnamefont
  {Vassilevich}},\ }\bibfield  {title} {\bibinfo {title} {{Heat kernel
  expansion: user's manual}},\ }\href
  {https://doi.org/10.1016/j.physrep.2003.09.002} {\bibfield  {journal}
  {\bibinfo  {journal} {Phys. Rep.}\ }\textbf {\bibinfo {volume} {388}},\
  \bibinfo {pages} {279} (\bibinfo {year} {2003})},\ \Eprint
  {https://arxiv.org/abs/0306138} {arXiv:0306138 [hep-th]} \BibitemShut
  {NoStop}%
\bibitem [{\citenamefont {'t~Hooft}\ and\ \citenamefont
  {Veltman}(1974)}]{tHooftVeltman}%
  \BibitemOpen
  \bibfield  {author} {\bibinfo {author} {\bibfnamefont {G.}~\bibnamefont
  {'t~Hooft}}\ and\ \bibinfo {author} {\bibfnamefont {M.}~\bibnamefont
  {Veltman}},\ }\bibfield  {title} {\bibinfo {title} {{One loop divergencies in
  the theory of gravitation}},\ }\href
  {https://doi.org/10.1142/9789814539395_0001} {\bibfield  {journal} {\bibinfo
  {journal} {Ann. Inst. Henri Poincare}\ ,\ \bibinfo {pages} {69}} (\bibinfo
  {year} {1974})}\BibitemShut {NoStop}%
\bibitem [{\citenamefont {Gibbons}(1979)}]{Gibbons}%
  \BibitemOpen
  \bibfield  {author} {\bibinfo {author} {\bibfnamefont {G.~W.}\ \bibnamefont
  {Gibbons}},\ }\bibfield  {title} {\bibinfo {title} {{Quantum field theory in
  curved spacetime}},\ }in\ \href@noop {} {\emph {\bibinfo {booktitle}
  {{General Relativity. An Einstein Centenary Survey}}}}\ (\bibinfo
  {publisher} {Cambridge University Press},\ \bibinfo {address} {Cambridge,
  England},\ \bibinfo {year} {1979})\ pp.\ \bibinfo {pages}
  {639--679}\BibitemShut {NoStop}%
\bibitem [{\citenamefont {Fradkin}\ and\ \citenamefont
  {Tseytlin}(1982)}]{Fradkin1982}%
  \BibitemOpen
  \bibfield  {author} {\bibinfo {author} {\bibfnamefont {E.~S.}\ \bibnamefont
  {Fradkin}}\ and\ \bibinfo {author} {\bibfnamefont {A.~A.}\ \bibnamefont
  {Tseytlin}},\ }\bibfield  {title} {\bibinfo {title} {{Renormalizable
  asymptotically free quantum theory of gravity}},\ }\href
  {https://doi.org/10.1016/0550-3213(82)90444-8} {\bibfield  {journal}
  {\bibinfo  {journal} {Nucl. Phys.}\ }\textbf {\bibinfo {volume} {B201}},\
  \bibinfo {pages} {469} (\bibinfo {year} {1982})}\BibitemShut {NoStop}%
\bibitem [{\citenamefont {Avramidy}\ and\ \citenamefont
  {Barvinsky}(1985)}]{AvramBarvinsky}%
  \BibitemOpen
  \bibfield  {author} {\bibinfo {author} {\bibfnamefont {I.~G.}\ \bibnamefont
  {Avramidy}}\ and\ \bibinfo {author} {\bibfnamefont {A.~O.}\ \bibnamefont
  {Barvinsky}},\ }\bibfield  {title} {\bibinfo {title} {{Asymptotic freedom in
  higher-derivative quantum gravity}},\ }\href
  {https://doi.org/10.1016/0370-2693(85)90248-5} {\bibfield  {journal}
  {\bibinfo  {journal} {Phys. Lett.}\ }\textbf {\bibinfo {volume} {B159}},\
  \bibinfo {pages} {269} (\bibinfo {year} {1985})}\BibitemShut {NoStop}%
\bibitem [{\citenamefont {Jack}\ and\ \citenamefont
  {Osborn}(1984)}]{JackOsborn}%
  \BibitemOpen
  \bibfield  {author} {\bibinfo {author} {\bibfnamefont {I.}~\bibnamefont
  {Jack}}\ and\ \bibinfo {author} {\bibfnamefont {H.}~\bibnamefont {Osborn}},\
  }\bibfield  {title} {\bibinfo {title} {{Background field calculations in
  curved spacetime (I). General formalism and application to scalar fields}},\
  }\href {https://doi.org/10.1016/0550-3213(84)90067-1} {\bibfield  {journal}
  {\bibinfo  {journal} {Nucl. Phys.}\ }\textbf {\bibinfo {volume} {B234}},\
  \bibinfo {pages} {331} (\bibinfo {year} {1984})}\BibitemShut {NoStop}%
\bibitem [{\citenamefont {Barvinsky}\ \emph {et~al.}(2021)\citenamefont
  {Barvinsky}, \citenamefont {Kurov},\ and\ \citenamefont
  {Sibiryakov}}]{Barvinsky2021ubv}%
  \BibitemOpen
  \bibfield  {author} {\bibinfo {author} {\bibfnamefont {A.~O.}\ \bibnamefont
  {Barvinsky}}, \bibinfo {author} {\bibfnamefont {A.~V.}\ \bibnamefont
  {Kurov}},\ and\ \bibinfo {author} {\bibfnamefont {S.~M.}\ \bibnamefont
  {Sibiryakov}},\ }\bibfield  {title} {\bibinfo {title} {{Beta functions of
  $(3+1)$-dimensional projectable Ho\v{r}ava gravity}},\ }\Eprint
  {https://arxiv.org/abs/2110.14688} {arXiv:2110.14688 [hep-th]}  (\bibinfo
  {year} {2021})\BibitemShut {NoStop}%
\bibitem [{\citenamefont {Barvinsky}\ \emph {et~al.}(2019)\citenamefont
  {Barvinsky}, \citenamefont {Pronin},\ and\ \citenamefont
  {Wachowski}}]{Wach2}%
  \BibitemOpen
  \bibfield  {author} {\bibinfo {author} {\bibfnamefont {A.~O.}\ \bibnamefont
  {Barvinsky}}, \bibinfo {author} {\bibfnamefont {P.~I.}\ \bibnamefont
  {Pronin}},\ and\ \bibinfo {author} {\bibfnamefont {W.}~\bibnamefont
  {Wachowski}},\ }\bibfield  {title} {\bibinfo {title} {{Heat kernel for
  higher-order differential operators and generalized exponential functions}},\
  }\href {https://doi.org/10.1103/PhysRevD.100.105004} {\bibfield  {journal}
  {\bibinfo  {journal} {Phys. Rev.}\ }\textbf {\bibinfo {volume} {D100}},\
  \bibinfo {pages} {105004} (\bibinfo {year} {2019})},\ \Eprint
  {https://arxiv.org/abs/1908.02161} {arXiv:1908.02161 [hep-th]} \BibitemShut
  {NoStop}%
\bibitem [{\citenamefont {Paneitz}(2008)}]{Paneitz}%
  \BibitemOpen
  \bibfield  {author} {\bibinfo {author} {\bibfnamefont {S.}~\bibnamefont
  {Paneitz}},\ }\bibfield  {title} {\bibinfo {title} {{A quartic conformally
  covariant differential operator for arbitrary pseudo-Riemannian manifolds
  (summary)}},\ }in\ \href {https://doi.org/10.3842/SIGMA.2008.036} {\emph
  {\bibinfo {booktitle} {{Symmetry, Integrability and Geometry: Methods ans
  Applications (SIGMA)}}}},\ Vol.~\bibinfo {volume} {4}\ (\bibinfo {year}
  {2008})\BibitemShut {NoStop}%
\bibitem [{\citenamefont {Branson}(1996)}]{Branson}%
  \BibitemOpen
  \bibfield  {author} {\bibinfo {author} {\bibfnamefont {T.~P.}\ \bibnamefont
  {Branson}},\ }\bibfield  {title} {\bibinfo {title} {{An anomaly associated
  with 4-dimensional quantum gravity}},\ }\href
  {https://doi.org/10.1007/BF02099450} {\bibfield  {journal} {\bibinfo
  {journal} {Commun. Math. Phys.}\ }\textbf {\bibinfo {volume} {178}},\
  \bibinfo {pages} {301} (\bibinfo {year} {1996})}\BibitemShut {NoStop}%
\bibitem [{\citenamefont {Erdmenger}(1997)}]{Erdmenger}%
  \BibitemOpen
  \bibfield  {author} {\bibinfo {author} {\bibfnamefont {J.}~\bibnamefont
  {Erdmenger}},\ }\bibfield  {title} {\bibinfo {title} {{Conformally covariant
  differential operators: properties and applications}},\ }\href
  {https://doi.org/10.1088/0264-9381/14/8/008} {\bibfield  {journal} {\bibinfo
  {journal} {Class. Quantum Grav.}\ }\textbf {\bibinfo {volume} {14}},\
  \bibinfo {pages} {2061} (\bibinfo {year} {1997})}\BibitemShut {NoStop}%
\bibitem [{\citenamefont {Maslov}\ and\ \citenamefont
  {Fedoriuk}(1981)}]{MaslovFedoriuk}%
  \BibitemOpen
  \bibfield  {author} {\bibinfo {author} {\bibfnamefont {V.~P.}\ \bibnamefont
  {Maslov}}\ and\ \bibinfo {author} {\bibfnamefont {M.~V.}\ \bibnamefont
  {Fedoriuk}},\ }\href@noop {} {\emph {\bibinfo {title} {{Semi-classical
  approximation in quantum mechanics}}}}\ (\bibinfo  {publisher} {D. Reidel
  Publishing Company},\ \bibinfo {address} {Dordrecht},\ \bibinfo {year}
  {1981})\BibitemShut {NoStop}%
\bibitem [{\citenamefont {Carinhas}\ and\ \citenamefont
  {Fulling}(1990)}]{Fulling}%
  \BibitemOpen
  \bibfield  {author} {\bibinfo {author} {\bibfnamefont {P.~A.}\ \bibnamefont
  {Carinhas}}\ and\ \bibinfo {author} {\bibfnamefont {S.~A.}\ \bibnamefont
  {Fulling}},\ }\bibfield  {title} {\bibinfo {title} {{Computational
  asymptotics of fourth-order operators}},\ }in\ \href@noop {} {\emph {\bibinfo
  {booktitle} {Asymptotic and computational analysis: conference in honor of
  Frank W.J. Olver's 65th birthday}}},\ \bibinfo {series and number} {Proc.
  Sympos. Pure Math.}\ (\bibinfo  {publisher} {MARCEL DEKKER, Inc.},\ \bibinfo
  {address} {New York},\ \bibinfo {year} {1990})\BibitemShut {NoStop}%
\bibitem [{\citenamefont {Gilkey}(2003)}]{Gilkey2003}%
  \BibitemOpen
  \bibfield  {author} {\bibinfo {author} {\bibfnamefont {P.~B.}\ \bibnamefont
  {Gilkey}},\ }\href@noop {} {\emph {\bibinfo {title} {{Asymptotic Formulae in
  Spectral Geometry}}}}\ (\bibinfo  {publisher} {Chapman and Hall/CRC},\
  \bibinfo {address} {Boca Raton, London, New York, Washington, DC},\ \bibinfo
  {year} {2003})\BibitemShut {NoStop}%
\bibitem [{\citenamefont {Gilkey}(1980)}]{Gilkey1980}%
  \BibitemOpen
  \bibfield  {author} {\bibinfo {author} {\bibfnamefont {P.~B.}\ \bibnamefont
  {Gilkey}},\ }\bibfield  {title} {\bibinfo {title} {{The spectral geometry of
  the higher order Laplacian}},\ }\href
  {https://doi.org/10.1215/S0012-7094-80-04731-6} {\bibfield  {journal}
  {\bibinfo  {journal} {Duke Math. J.}\ }\textbf {\bibinfo {volume} {47}},\
  \bibinfo {pages} {511} (\bibinfo {year} {1980})}\BibitemShut {NoStop}%
\bibitem [{\citenamefont {Fegan}\ and\ \citenamefont
  {Gilkey}(1985)}]{GilkeyFegan}%
  \BibitemOpen
  \bibfield  {author} {\bibinfo {author} {\bibfnamefont {H.~D.}\ \bibnamefont
  {Fegan}}\ and\ \bibinfo {author} {\bibfnamefont {P.~B.}\ \bibnamefont
  {Gilkey}},\ }\bibfield  {title} {\bibinfo {title} {{Invariants of the heat
  equation}},\ }\href {https://doi.org/10.0000/msp.org/pjm/1985/117-2/p03}
  {\bibfield  {journal} {\bibinfo  {journal} {Pac. J. Math.}\ }\textbf
  {\bibinfo {volume} {117}},\ \bibinfo {pages} {233} (\bibinfo {year}
  {1985})}\BibitemShut {NoStop}%
\bibitem [{\citenamefont {Gilkey}\ \emph {et~al.}(1991)\citenamefont {Gilkey},
  \citenamefont {Branson},\ and\ \citenamefont
  {Fulling}}]{GilkeyBransonFulling}%
  \BibitemOpen
  \bibfield  {author} {\bibinfo {author} {\bibfnamefont {P.~B.}\ \bibnamefont
  {Gilkey}}, \bibinfo {author} {\bibfnamefont {T.~P.}\ \bibnamefont
  {Branson}},\ and\ \bibinfo {author} {\bibfnamefont {S.~A.}\ \bibnamefont
  {Fulling}},\ }\bibfield  {title} {\bibinfo {title} {{Heat equation
  asymptotics of ``nonminimal'' operators on differential forms}},\ }\href
  {https://doi.org/10.1063/1.529179} {\bibfield  {journal} {\bibinfo  {journal}
  {J. Math. Phys. (N.Y.)}\ }\textbf {\bibinfo {volume} {32}},\ \bibinfo {pages}
  {2089} (\bibinfo {year} {1991})}\BibitemShut {NoStop}%
\bibitem [{\citenamefont {Widom}(1978)}]{Widom1}%
  \BibitemOpen
  \bibfield  {author} {\bibinfo {author} {\bibfnamefont {H.}~\bibnamefont
  {Widom}},\ }\bibfield  {title} {\bibinfo {title} {{Families of
  pseudodifferential operators}},\ }in\ \href@noop {} {\emph {\bibinfo
  {booktitle} {Topics in functional analysis: Essays dedicated to {M.G.~Krein}
  on the occasion of his 70th birthday}}},\ \bibinfo {series} {Adv. in Math.
  Suppl. Stud.}, Vol.~\bibinfo {volume} {3}\ (\bibinfo  {publisher} {Academic
  press},\ \bibinfo {address} {New York, San Francisco, London},\ \bibinfo
  {year} {1978})\ pp.\ \bibinfo {pages} {345--395}\BibitemShut {NoStop}%
\bibitem [{\citenamefont {Widom}(1979)}]{Widom2}%
  \BibitemOpen
  \bibfield  {author} {\bibinfo {author} {\bibfnamefont {H.}~\bibnamefont
  {Widom}},\ }\bibfield  {title} {\bibinfo {title} {{Szeg\"o’s theorem and a
  complete symbolic calculus for pseudodifferential operators}},\ }in\
  \href@noop {} {\emph {\bibinfo {booktitle} {Seminar on Singularities of
  Solutions of Linear Partial Differential Equations}}}\ (\bibinfo  {publisher}
  {Princeton Univ. Press},\ \bibinfo {address} {Princeton},\ \bibinfo {year}
  {1979})\ pp.\ \bibinfo {pages} {261--283}\BibitemShut {NoStop}%
\bibitem [{\citenamefont {Widom}(1980)}]{Widom3}%
  \BibitemOpen
  \bibfield  {author} {\bibinfo {author} {\bibfnamefont {H.}~\bibnamefont
  {Widom}},\ }\bibfield  {title} {\bibinfo {title} {{A complete symbolic
  calculus for pseudodifferential operators}},\ }\href@noop {} {\bibfield
  {journal} {\bibinfo  {journal} {Bulletin des Sciences Math{\'e}matiques}\
  }\textbf {\bibinfo {volume} {104}},\ \bibinfo {pages} {19} (\bibinfo {year}
  {1980})}\BibitemShut {NoStop}%
\bibitem [{\citenamefont {Gusynin}(1989)}]{Gusynin1989}%
  \BibitemOpen
  \bibfield  {author} {\bibinfo {author} {\bibfnamefont {V.~P.}\ \bibnamefont
  {Gusynin}},\ }\bibfield  {title} {\bibinfo {title} {{New algorithm for
  computing the coefficients in the heat kernel expansion}},\ }\href
  {https://doi.org/10.1016/0370-2693(89)90811-3} {\bibfield  {journal}
  {\bibinfo  {journal} {Phys. Lett.}\ }\textbf {\bibinfo {volume} {B225}},\
  \bibinfo {pages} {233} (\bibinfo {year} {1989})}\BibitemShut {NoStop}%
\bibitem [{\citenamefont {Gusynin}(1990)}]{Gusynin1990}%
  \BibitemOpen
  \bibfield  {author} {\bibinfo {author} {\bibfnamefont {V.~P.}\ \bibnamefont
  {Gusynin}},\ }\bibfield  {title} {\bibinfo {title} {{Seeley--Gilkey
  coefficients for fourth-order operators on a Riemannian manifold}},\ }\href
  {https://doi.org/10.1016/0550-3213(90)90233-4} {\bibfield  {journal}
  {\bibinfo  {journal} {Nucl. Phys.}\ }\textbf {\bibinfo {volume} {B333}},\
  \bibinfo {pages} {296} (\bibinfo {year} {1990})}\BibitemShut {NoStop}%
\bibitem [{\citenamefont {Gusynin}(1991)}]{Gusynin1991}%
  \BibitemOpen
  \bibfield  {author} {\bibinfo {author} {\bibfnamefont {V.~P.}\ \bibnamefont
  {Gusynin}},\ }\bibfield  {title} {\bibinfo {title} {{Asymptotics of the heat
  kernel for nonminimal differential operators}},\ }\href
  {https://doi.org/10.1007/BF01067283} {\bibfield  {journal} {\bibinfo
  {journal} {Ukr. Math. J.}\ }\textbf {\bibinfo {volume} {43}},\ \bibinfo
  {pages} {1432} (\bibinfo {year} {1991})}\BibitemShut {NoStop}%
\bibitem [{\citenamefont {Gusynin}\ and\ \citenamefont
  {Gorbar}(1991)}]{GusyninGorbar}%
  \BibitemOpen
  \bibfield  {author} {\bibinfo {author} {\bibfnamefont {V.~P.}\ \bibnamefont
  {Gusynin}}\ and\ \bibinfo {author} {\bibfnamefont {E.~V.}\ \bibnamefont
  {Gorbar}},\ }\bibfield  {title} {\bibinfo {title} {{Local heat kernel
  asymptotics for nonminimal differential operators}},\ }\href
  {https://doi.org/10.1016/0370-2693(91)91534-3} {\bibfield  {journal}
  {\bibinfo  {journal} {Phys. Lett.}\ }\textbf {\bibinfo {volume} {B270}},\
  \bibinfo {pages} {29} (\bibinfo {year} {1991})}\BibitemShut {NoStop}%
\bibitem [{\citenamefont {Gusynin}\ \emph {et~al.}(1991)\citenamefont
  {Gusynin}, \citenamefont {Gorbar},\ and\ \citenamefont
  {Romankov}}]{GusyninGorbarRomankov}%
  \BibitemOpen
  \bibfield  {author} {\bibinfo {author} {\bibfnamefont {V.~P.}\ \bibnamefont
  {Gusynin}}, \bibinfo {author} {\bibfnamefont {E.~V.}\ \bibnamefont
  {Gorbar}},\ and\ \bibinfo {author} {\bibfnamefont {V.~V.}\ \bibnamefont
  {Romankov}},\ }\bibfield  {title} {\bibinfo {title} {{Heat kernel expansion
  for nonminimal differential operations and manifolds with torsion}},\ }\href
  {https://doi.org/10.1016/0550-3213(91)90568-i} {\bibfield  {journal}
  {\bibinfo  {journal} {Nucl. Phys.}\ }\textbf {\bibinfo {volume} {B362}},\
  \bibinfo {pages} {449} (\bibinfo {year} {1991})}\BibitemShut {NoStop}%
\bibitem [{\citenamefont {Gorbar}(1997)}]{Gorbar}%
  \BibitemOpen
  \bibfield  {author} {\bibinfo {author} {\bibfnamefont {E.~V.}\ \bibnamefont
  {Gorbar}},\ }\bibfield  {title} {\bibinfo {title} {{Heat kernel expansion for
  operators containing a root of the Laplace operator}},\ }\href
  {https://doi.org/10.1063/1.531823} {\bibfield  {journal} {\bibinfo  {journal}
  {J. Math. Phys. (N.Y.)}\ }\textbf {\bibinfo {volume} {38}},\ \bibinfo {pages}
  {1692} (\bibinfo {year} {1997})},\ \Eprint {https://arxiv.org/abs/9602018}
  {arXiv:9602018 [hep-th]} \BibitemShut {NoStop}%
\bibitem [{\citenamefont {Gilkey}\ and\ \citenamefont
  {Grubb}(1998)}]{GilkeyGrubb}%
  \BibitemOpen
  \bibfield  {author} {\bibinfo {author} {\bibfnamefont {P.~B.}\ \bibnamefont
  {Gilkey}}\ and\ \bibinfo {author} {\bibfnamefont {G.}~\bibnamefont {Grubb}},\
  }\bibfield  {title} {\bibinfo {title} {{Logarithmic terms in asymptotic
  expansions of heat operator traces}},\ }\href
  {https://doi.org/10.1080/03605309808821365} {\bibfield  {journal} {\bibinfo
  {journal} {Commun. Partial Differ. Equations}\ }\textbf {\bibinfo {volume}
  {23}},\ \bibinfo {pages} {777} (\bibinfo {year} {1998})}\BibitemShut
  {NoStop}%
\bibitem [{\citenamefont {B{\"a}r}\ and\ \citenamefont
  {Moroianu}(2003)}]{Bar2003}%
  \BibitemOpen
  \bibfield  {author} {\bibinfo {author} {\bibfnamefont {C.}~\bibnamefont
  {B{\"a}r}}\ and\ \bibinfo {author} {\bibfnamefont {S.}~\bibnamefont
  {Moroianu}},\ }\bibfield  {title} {\bibinfo {title} {{Heat kernel asymptotics
  for roots of generalized Laplacians}},\ }\href
  {https://doi.org/10.1142/S0129167X03001788} {\bibfield  {journal} {\bibinfo
  {journal} {Int. J. Math.}\ }\textbf {\bibinfo {volume} {14}},\ \bibinfo
  {pages} {397} (\bibinfo {year} {2003})}\BibitemShut {NoStop}%
\bibitem [{\citenamefont {Ivanov}(2019)}]{Ivanov2019}%
  \BibitemOpen
  \bibfield  {author} {\bibinfo {author} {\bibfnamefont {A.~V.}\ \bibnamefont
  {Ivanov}},\ }\bibfield  {title} {\bibinfo {title} {{Diagram Technique for the
  Heat Kernel of the Covariant Laplace Operator}},\ }\href
  {https://doi.org/10.1134/S0040577919010070} {\bibfield  {journal} {\bibinfo
  {journal} {Theor. Math. Phys.}\ }\textbf {\bibinfo {volume} {198}},\ \bibinfo
  {pages} {100} (\bibinfo {year} {2019})},\ \Eprint
  {https://arxiv.org/abs/1905.05455} {arXiv:1905.05455 [hep-th]} \BibitemShut
  {NoStop}%
\bibitem [{\citenamefont {Ivanov}\ and\ \citenamefont
  {Kharuk}(2020)}]{Ivanov2019uxu}%
  \BibitemOpen
  \bibfield  {author} {\bibinfo {author} {\bibfnamefont {A.~V.}\ \bibnamefont
  {Ivanov}}\ and\ \bibinfo {author} {\bibfnamefont {N.~V.}\ \bibnamefont
  {Kharuk}},\ }\bibfield  {title} {\bibinfo {title} {{Heat kernel: Proper-time
  method, Fock\textendash{}Schwinger gauge, path integral, and Wilson line}},\
  }\href {https://doi.org/10.1134/S0040577920110057} {\bibfield  {journal}
  {\bibinfo  {journal} {Theor. Math. Phys.}\ }\textbf {\bibinfo {volume}
  {205}},\ \bibinfo {pages} {1456} (\bibinfo {year} {2020})},\ \Eprint
  {https://arxiv.org/abs/1906.04019} {arXiv:1906.04019 [hep-th]} \BibitemShut
  {NoStop}%
\bibitem [{\citenamefont {Nesterov}\ and\ \citenamefont
  {Solodukhin}(2011)}]{Solodukhin2011}%
  \BibitemOpen
  \bibfield  {author} {\bibinfo {author} {\bibfnamefont {D.}~\bibnamefont
  {Nesterov}}\ and\ \bibinfo {author} {\bibfnamefont {S.~N.}\ \bibnamefont
  {Solodukhin}},\ }\bibfield  {title} {\bibinfo {title} {{Gravitational
  effective action and entanglement entropy in UV modified theories with and
  without Lorentz symmetry}},\ }\href
  {https://doi.org/10.1016/j.nuclphysb.2010.08.006} {\bibfield  {journal}
  {\bibinfo  {journal} {Nucl. Phys. B}\ }\textbf {\bibinfo {volume} {842}},\
  \bibinfo {pages} {141} (\bibinfo {year} {2011})},\ \Eprint
  {https://arxiv.org/abs/1007.1246} {arXiv:1007.1246 [hep-th]} \BibitemShut
  {NoStop}%
\bibitem [{\citenamefont {Barvinsky}\ \emph {et~al.}(2017)\citenamefont
  {Barvinsky}, \citenamefont {Blas}, \citenamefont {Herrero-Valea},
  \citenamefont {Sibiryakov},\ and\ \citenamefont {Steinwachs}}]{Barvinsky172}%
  \BibitemOpen
  \bibfield  {author} {\bibinfo {author} {\bibfnamefont {A.~O.}\ \bibnamefont
  {Barvinsky}}, \bibinfo {author} {\bibfnamefont {D.}~\bibnamefont {Blas}},
  \bibinfo {author} {\bibfnamefont {M.}~\bibnamefont {Herrero-Valea}}, \bibinfo
  {author} {\bibfnamefont {S.~M.}\ \bibnamefont {Sibiryakov}},\ and\ \bibinfo
  {author} {\bibfnamefont {C.~F.}\ \bibnamefont {Steinwachs}},\ }\bibfield
  {title} {\bibinfo {title} {{Ho\v{r}ava gravity is asymptotically free in
  $2+1$ dimensions}},\ }\href {https://doi.org/10.1103/PhysRevLett.119.211301}
  {\bibfield  {journal} {\bibinfo  {journal} {Phys. Rev. Lett.}\ }\textbf
  {\bibinfo {volume} {119}},\ \bibinfo {pages} {211301} (\bibinfo {year}
  {2017})},\ \Eprint {https://arxiv.org/abs/1706.06809} {arXiv:1706.06809
  [hep-th]} \BibitemShut {NoStop}%
\bibitem [{\citenamefont {Groh}\ \emph {et~al.}(2011)\citenamefont {Groh},
  \citenamefont {Saueressig},\ and\ \citenamefont {Zanusso}}]{Saueressig_etal}%
  \BibitemOpen
  \bibfield  {author} {\bibinfo {author} {\bibfnamefont {K.}~\bibnamefont
  {Groh}}, \bibinfo {author} {\bibfnamefont {F.}~\bibnamefont {Saueressig}},\
  and\ \bibinfo {author} {\bibfnamefont {O.}~\bibnamefont {Zanusso}},\
  }\bibfield  {title} {\bibinfo {title} {{Off-diagonal heat-kernel expansion
  and its application to fields with differential constraints}},\ }\href@noop
  {} {\  (\bibinfo {year} {2011})},\ \Eprint {https://arxiv.org/abs/1112.4856}
  {arXiv:1112.4856 [math-ph]} \BibitemShut {NoStop}%
\bibitem [{\citenamefont {Grosvenor}\ \emph {et~al.}(2021)\citenamefont
  {Grosvenor}, \citenamefont {Melby-Thompson},\ and\ \citenamefont
  {Yan}}]{Grosvenor2021zvq}%
  \BibitemOpen
  \bibfield  {author} {\bibinfo {author} {\bibfnamefont {K.~T.}\ \bibnamefont
  {Grosvenor}}, \bibinfo {author} {\bibfnamefont {C.}~\bibnamefont
  {Melby-Thompson}},\ and\ \bibinfo {author} {\bibfnamefont {Z.}~\bibnamefont
  {Yan}},\ }\bibfield  {title} {\bibinfo {title} {{New Heat Kernel Method in
  Lifshitz Theories}},\ }\href {https://doi.org/10.1007/JHEP04(2021)178}
  {\bibfield  {journal} {\bibinfo  {journal} {JHEP}\ }\textbf {\bibinfo
  {volume} {04}},\ \bibinfo {pages} {178}},\ \Eprint
  {https://arxiv.org/abs/2101.03177} {arXiv:2101.03177 [hep-th]} \BibitemShut
  {NoStop}%
\bibitem [{\citenamefont {Barvinsky}\ and\ \citenamefont
  {Wachowski}()}]{Work_in_progress}%
  \BibitemOpen
  \bibfield  {author} {\bibinfo {author} {\bibfnamefont {A.~O.}\ \bibnamefont
  {Barvinsky}}\ and\ \bibinfo {author} {\bibfnamefont {W.}~\bibnamefont
  {Wachowski}},\ }\bibfield  {title} {\bibinfo {title} {{Universal functional
  traces method for heat kenel of generic minimal operators}},\ }\bibinfo
  {note} {work in progress}\BibitemShut {NoStop}%
\bibitem [{\citenamefont {Wachowski}\ and\ \citenamefont
  {Pronin}(2019)}]{Wach}%
  \BibitemOpen
  \bibfield  {author} {\bibinfo {author} {\bibfnamefont {W.~N.}\ \bibnamefont
  {Wachowski}}\ and\ \bibinfo {author} {\bibfnamefont {P.~I.}\ \bibnamefont
  {Pronin}},\ }\bibfield  {title} {\bibinfo {title} {{The evolution function of
  the operator $-(-{\Delta})^\nu$}},\ }\href@noop {} {\bibfield  {journal}
  {\bibinfo  {journal} {Moscow University Physics Bulletin}\ }\textbf {\bibinfo
  {volume} {74}},\ \bibinfo {pages} {17} (\bibinfo {year} {2019})}\BibitemShut
  {NoStop}%
\end{thebibliography}%

\end{document}